\documentclass[a4paper,12pt]{article}

\usepackage{amsmath,amsthm,amsbsy,amssymb,amsfonts,graphicx,mathrsfs,bm,accents,cite,xcolor,color}
\usepackage[breaklinks=true]{hyperref}

\makeatletter
\catcode`\@=11
\@addtoreset{equation}{section}

\makeatother

\renewcommand{\thefootnote}{\arabic{footnote}}

\newcommand{\Exp}[1]{\operatorname{e}^{#1}}
\newcommand{\abs}[1]{\lvert{#1} \rvert}
\newcommand{\rmd}{{\mathrm{d}}}
\newcommand{\dd}{\rmd}
\newcommand{\nn}{\nonumber}
\newcommand{\no}{\nonumber}
\newcommand{\Lie}{\pounds}
\newcommand{\gLie}{\hat{\pounds}}

\newcommand{\cF}{\mathcal F}
\newcommand{\cH}{\mathcal H}

\newcommand{\cK}{\mathcal K}

\newcommand{\cO}{\mathcal O}

\newcommand{\sfA}{\mathsf A}
\newcommand{\sfC}{\mathsf C}
\newcommand{\sfF}{\mathsf F}

\newcommand{\rmT}{\mathrm{T}}

\newcommand{\GL}{\mathrm{GL}}
\newcommand{\OO}{\mathrm{O}}

\newcommand{\bB}{{\bm B}}
\newcommand{\bbeta}{{\bm\beta}}
\newcommand{\OG}{G}
\newcommand{\CG}{g}

\newcommand*{\AdS}[1]{\ensuremath{\text{AdS}_{#1}}}
\newcommand{\rmS}{{\textrm{S}}}

\allowdisplaybreaks[3]

\begin{document}

\begin{titlepage}
\renewcommand{\thefootnote}{\fnsymbol{footnote}}

\begin{flushright}
\parbox{4cm}{
\hfill
KUNS-2682

\hfill
YITP-17-110
}
\hspace*{1.5cm}
\end{flushright}

\vspace*{1cm}

\begin{center}
{\Large\textbf{
$T$-folds from Yang-Baxter deformations
}}%
\end{center}
\vspace{1.0cm}

\centerline{
{\large Jos\'e J.~Fern\'andez-Melgarejo$^{a,b}$},%
\footnote{E-mail address: \texttt{josejuan@yukawa.kyoto-u.ac.jp}}
{\large Jun-ichi Sakamoto$^c$},%
\footnote{E-mail address: \texttt{sakajun@gauge.scphys.kyoto-u.ac.jp}}}
\centerline{
{\large Yuho Sakatani$^{d,e}$},%
\footnote{E-mail address: \texttt{yuho@koto.kpu-m.ac.jp}}
and {\large Kentaroh Yoshida$^c$}%
\footnote{E-mail address: \texttt{kyoshida@gauge.scphys.kyoto-u.ac.jp}}
}

\vspace{0.2cm}

\begin{center}
${}^a${\it Yukawa Institute for Theoretical Physics, Kyoto University}

\vspace*{1mm}

${}^b${\it Departamento de F\'isica, Universidad de Murcia}

\vspace*{1mm}

${}^c${\it Department of Physics, Kyoto University}

\vspace*{1mm}

${}^d${\it Department of Physics, Kyoto Prefectural University of Medicine}

\vspace*{1mm}

${}^e${\it Fields, Gravity \& Strings, CTPU, Institute for Basic Sciences}
\end{center}

\vspace*{1mm}

\begin{abstract}
Yang-Baxter (YB) deformations of type IIB string theory have been well studied from the viewpoint of classical integrability. 
Most of the works, however, are focused upon the local structure 
of the deformed geometries and the global structure still remains unclear. In this work, we reveal a non-geometric aspect of YB-deformed 
backgrounds as $T$-fold by explicitly showing the associated $\OO(D,D;\mathbb{Z})$ $T$-duality monodromy. In particular, the appearance of an extra vector field in the generalized supergravity equations (GSE) leads to the non-geometric $Q$-flux. 
In addition, we study a particular solution of GSE that is obtained by a non-Abelian $T$-duality but cannot be expressed as a homogeneous YB deformation, 
and show that it can also be regarded as a $T$-fold. This result indicates that solutions of GSE should be non-geometric quite in general beyond the YB deformation. 
\end{abstract}

\thispagestyle{empty}
\end{titlepage}

\setcounter{footnote}{0}

\tableofcontents

\section{Introduction}

A prototypical example of the AdS/CFT correspondence \cite{Maldacena:1997re} is a conjectured equivalence between a type IIB superstring theory 
on $\AdS5\times\rmS^5$ and the four-dimensional $\mathcal{N}=4$ SU$(N)$ super Yang-Mills theory in the large $N$ limit. 
Nowadays, it is well recognized that an integrable structure underlies this correspondence (for a comprehensive review, see \cite{Beisert:2010jr}). 
In particular, the associated supergeometry is represented by a supercoset \cite{Metsaev:1998it},
\begin{align}
 \frac{\text{PSU}(2,2|4)}{\text{SO}(1,4) \times \text{SO}(5)}\,,
\label{eq:supercoset}
\end{align}
and the $\mathbb{Z}_4$-grading of it ensures the classical integrability of the supercoset sigma model \cite{Bena:2003wd}. 

\medskip 

A renewed interest for this integrable structure appeared from the development of a systematic scheme of integrable deformation 
called the Yang-Baxter (YB) deformation \cite{Klimcik:2002zj,Klimcik:2008eq,Klimcik:2014bta,Delduc:2013fga,Matsumoto:2015jja}. 
In fact, an application of this scheme to type IIB superstring on $\AdS5\times \rmS^5$ 
\cite{Delduc:2013qra,Delduc:2014kha,Kawaguchi:2014qwa} opened up a lot of new perspectives and directions 
including intriguing relations among YB deformations and non-commutative gauge theories, non-Abelian $T$-duality 
\cite{Fridling:1983ha,Fradkin:1984ai,delaOssa:1992vci,Gasperini:1993nz,Giveon:1993ai,Alvarez:1994np,
Elitzur:1994ri,Sfetsos:2010uq,Lozano:2011kb,Itsios:2013wd}, 
and manifestly $T$-/$U$-duality covariant formulations, which have been discovered in 
\cite{Matsumoto:2014gwa,vanTongeren:2016eeb,Araujo:2017jkb,Araujo:2017jap}, 
\cite{Hoare:2016wsk,Borsato:2016pas,Hoare:2016wca}, 
and \cite{Sakatani:2016fvh,Baguet:2016prz,Sakamoto:2017wor,Sakamoto:2017cpu}, respectively. 
Our concern here is to delve deeper into the relation to a manifestly $T$-duality covariant 
formulation called Double Field Theory (DFT) \cite{Siegel:1993xq,Siegel:1993th,Siegel:1993bj,Hull:2009mi,Hohm:2010pp} with particular emphasis 
on non-geometric aspects (\emph{i.e.}, the global nature) of YB-deformed backgrounds. 

\medskip

The original application of YB deformations to string theory is the standard $q$-deformation of type IIB string on $\AdS5\times\rmS^5$ \cite{Delduc:2013qra,Delduc:2014kha} with a classical $r$-matrix of Drinfeld-Jimbo type \cite{Drinfeld:1985rx,Jimbo:1985zk}. 
The metric and Neveu-Schwarz--Neveu-Schwarz (NS-NS) 2-form of the deformed background were computed in \cite{Arutyunov:2013ega} 
by performing coset construction for the bosonic group elements. 
Note here that the deformed background is called in several ways, as the $q$-deformed $\AdS5\times\rmS^5$\,, the $\eta$-deformed $\AdS5\times \rmS^5$\,, or the ABF background, 
but all of them are identical\footnote{It should be remarked that the $\eta$-deformed $\AdS5\times\rmS^5$ 
means all of the YB deformations of $\AdS5\times\rmS^5$ in some literature, but here we will not follow that convention.}. 
The supercoset construction for the $q$-deformed case was worked out in 
\cite{Arutyunov:2015qva}. While the $q$-deformation is based on the modified classical 
Yang-Baxter equation (mCYBE), one may consider another category based on 
the homogeneous CYBE \cite{Matsumoto:2015jja,Kawaguchi:2014qwa}, for which a large number of works 
\cite{Matsumoto:2014nra,Matsumoto:2014gwa,Matsumoto:2015uja,Kawaguchi:2014fca,Matsumoto:2014ubv,Matsumoto:2014cja,Matsumoto:2016lnr,vanTongeren:2015soa,vanTongeren:2015uha,Kameyama:2015ufa,Kyono:2016jqy,Hoare:2016hwh,Orlando:2016qqu,Borsato:2016ose,Osten:2016dvf,vanTongeren:2016eeb,Hoare:2016wca,Araujo:2017jkb} 
have been done and led to various backgrounds 
including well-known examples such as Lunin-Maldacena-Frolov backgrounds \cite{Lunin:2005jy,Frolov:2005dj}, 
gravity duals of non-commutative gauge theories \cite{Hashimoto:1999ut,Maldacena:1999mh} 
and Schr\"odinger spacetimes \cite{Herzog:2008wg,Maldacena:2008wh,Adams:2008wt}. 

\medskip 

Remarkably, the full $q$-deformed background of \cite{Arutyunov:2015qva}, 
which includes the Ramond--Ramond (R-R) fluxes and dilaton, 
is not a solution of type IIB supergravity. 
Afterward, it was shown that the background should satisfy 
the generalized supergravity equations of motion (GSE) \cite{Arutyunov:2015mqj}. 
At that moment, the GSE seemed to be an artifice invented 
so as to support the $q$-deformed background as a solution. 
However, after that, in the ground-breaking paper by Tseytlin and Wulff \cite{Wulff:2016tju}, 
this GSE has been reproduced 
by solving the kappa-symmetry constraints of the Green-Schwarz type IIB string theory 
on an arbitrary background. 
Therefore, the GSE has now been established on the fairly fundamental ground. 

\medskip 

For the homogeneous CYBE case \cite{Kawaguchi:2014qwa}, there exists a significant criterion 
to identify whether a YB-deformed background is a solution of type IIB supergravity or GSE, 
before deriving the concrete expression of the resulting deformed background, namely, 
at the level of classical $r$-matrix. It is called the unimodularity condition \cite{Borsato:2016ose}. 
When this condition is satisfied, the deformed background is a solution of type IIB supergravity, 
but if not, the background is a solution of the GSE. 
Various solutions of GSE have been obtained from YB deformations 
with non-unimodular classical $r$-matrices \cite{Kyono:2016jqy,Orlando:2016qqu}. As it has been shown 
in \cite{Matsumoto:2014gwa,vanTongeren:2016eeb,Araujo:2017jkb,Araujo:2017jap}, 
classical $r$-matrices, which characterize homogeneous YB deformations of $\AdS5$ geometry, 
are closely related to non-commutative parameters 
in the dual open-string description and, as pointed out in \cite{Sakamoto:2017cpu}, 
they are nothing but $\beta$-fields \cite{Duff:1989tf}. 
In terms of the $\beta$-field, the non-unimodularity is measured as 
\cite{Araujo:2017jkb,Araujo:2017jap,Sakamoto:2017cpu}
\begin{align}
 \frac{1}{\sqrt{\abs{\OG}}}\,\partial_m(\sqrt{\abs{\OG}}\,\beta^{nm}) \neq 0 \,,
\end{align}
where $\OG_{mn}$ is the so-called the open-string metric that will be defined later. 
The quantity on the left-hand side is basically the trace of a non-geometric $Q$-flux, 
and this result indicates that YB deformations with non-unimodular $r$-matrices lead to 
non-geometric backgrounds. 

\medskip 

In this paper, we will concentrate on YB deformations of Minkowski and 
$\AdS5\times \rmS^5$ backgrounds, and find that the deformed backgrounds we consider here 
belong to a specific class of non-geometric backgrounds, called $T$-folds \cite{Hull:2004in}. 
As far as we know, the YB-deformed backgrounds have not been recognized as $T$-folds so far, 
hence this is the first work that clearly states the relation between YB-deformed backgrounds 
and $T$-folds. Moreover, it is worth noting that our examples have an intriguing feature that 
the R-R fields are also twisted by the $T$-duality monodromy, in comparison to 
the well-known $T$-folds which include no R-R fields.

\medskip 

This paper is organized as follows. 
Section \ref{sec:T-fold} provides a brief review of $T$-folds, including two examples 
that are well-known in the literature. 
In Section \ref{sec:non-geometry}, we consider GSE solutions which can be realized 
as YB deformations of Minkowski spacetime and $\AdS5\times \rmS^5$, 
and argue that these deformed backgrounds are regarded as $T$-folds. 
In addition, we study a solution of GSE that is obtained by a non-Abelian $T$-duality 
but not as a YB deformation, and show that this can also be regarded as a $T$-fold. 
Section \ref{sec:discussion} is devoted to conclusions and discussions. 
In Appendix \ref{sec:Penrose-limit}, we discuss how to generalize the Penrose limit 
\cite{Penrose,Gueven:2000ru} so as to produce various GSE solutions. 
To be pedagogical, Appendix \ref{sec:Penrose-Ads5} is devoted 
to a review of Penrose limit of Poincar\'e $\AdS5$. 
In Appendix \ref{sec:Penrose-YB-deform}, we discuss the modified Penrose limit 
with a rescaling of the deformation parameter. 
Then, we apply it to YB-deformed background and reproduce 
the deformed Minkowski backgrounds discussed in Section \ref{sec:non-geometry-Minkowski}.

\section{A brief review of $T$-folds}
\label{sec:T-fold}

In this section, let us explain what is $T$-fold. A $T$-fold is supposed 
to be a generalization of the usual manifold. 
It locally looks like a Riemannian manifold, but which is glued together not just 
by diffeomorphisms but also by $T$-duality. 
It plays a significant role in studying non-geometric fluxes 
beyond the effective supergravity description. As illustrative examples, 
we revisit two well-known cases in the literature, corresponding to 
a chain of duality transformations \cite{Kachru:2002sk,Shelton:2005cf} 
and to the codimension-1 $5_2^2$-brane solution \cite{Hassler:2013wsa}. 

\medskip 

It is conjectured that string theories are related by some discrete dualities. 
One thing that can occur is that, by duality transformations, a flux configuration transforms 
into a non-geometric flux configuration, which means that it cannot be realized in terms of 
the usual fields in 10/11-dimensional supergravities. Therefore, dualities suggest that 
we need to go beyond the usual geometric isometries to fully understand 
the arena of flux compactifications.

\medskip 

For the case of $T$-duality, one proposal to address this problem is the so-called doubled formalism. 
This construction consists of a manifold in which all the local patches are geometric. 
However, the transition functions that are needed to glue these patches not only include usual diffeomorphisms and gauge transformations, but also $T$-duality transformations.

\medskip 

$T$-fold backgrounds are formulated in an enlarged space with a $T^n\times \tilde T^n$ fibration. The tangent space is the doubled torus $T^n\times \tilde T^n$ and is described by a set of coordinates $Y^M=(y^m, y_m)$ which transforms in the fundamental representation of $\OO(n,n)$. 
The physical internal space arises as a particular choice of a subspace of the double torus, $T^n_{\text{phys}}\subset T^n\times \tilde T^n$. 
Then $T$-duality transformations $\OO(n,n;\mathbb{Z})$ act by changing the physical subspace $T^n_{\text{phys}}$ to a different subspace of the enlarged $T^n \times \tilde T^n$. 
For a geometric background, we have a spacetime which is a geometric bundle, $T^n_{\text{phys}}=T^n$.\footnote{We can also have $T^n_{\text{phys}}=\tilde T^n$, which corresponds to a dual geometric description. 
} 
Nevertheless non-geometric backgrounds do not fit together to form a conventional manifold. 
That is to say, despite of they are locally well-defined, their global description is not valid. 
Instead, they are globally well-defined as $T$-folds.

\medskip 

This formulation is manifestly invariant under the $T$-duality group $O(n,n;\mathbb{Z})$. 
However, to make contact with the conventional formulation, one needs to choose a polarization, 
\emph{i.e.}, a particular choice of $T^n_{\text{phys}}\subset T^n\times \tilde T^n$. 
This means that we have to break the $\OO(n,n;\mathbb{Z})$ and pick $n$ coordinates out of 
the $2n$ coordinates $(y^m, \tilde y_m)$. Then, $T$-duality transformations allow to identify 
the backgrounds that belong to the same physical configuration or duality orbit and just differ 
on a choice of polarization\footnote{These orbits have been determined in terms of a classification of gauged supergravities in \cite{Dibitetto:2012rk}.}.

\medskip 

Due to the $\OO(n,n)$ symmetry, it is convenient to introduce the generalized metric 
$\cH_{MN}$ on the double torus,
\begin{align}
\begin{split}
 &\cH \equiv (\cH_{MN}) \equiv \Exp{-\bB^\rmT}\hat{\cH}\Exp{-\bB} 
 = \begin{pmatrix} (\CG-B\,\CG^{-1}\,B)_{mn} & B_{mk}\,\CG^{kn} \\
 -\CG^{mk}\,B_{kn} & \CG^{mn}
 \end{pmatrix} \,,
\\
 &\hat{\cH}\equiv \begin{pmatrix}
 \CG_{mn} & 0 \\
 0 & \CG^{mn}
 \end{pmatrix}\,,\quad
 (\bB^M{}_N) \equiv \begin{pmatrix}
 0 & 0 \\
 B_{mn} & 0
 \end{pmatrix}\,,\quad 
 \Exp{\bB} = \begin{pmatrix}
 \delta^m_n & 0 \\
 B_{mn} & \delta_m^n
 \end{pmatrix} \,, 
\end{split}
\label{eq:Hmetric}
\end{align}
where $g_{mn}$ and $B_{mn}$ are the internal components of the metric and 
the Kalb-Ramond 2-form, respectively. 
As $\cH\in \OO(n,n)$, the non-linear transformations of the $T$-duality group 
are covariantly realized as
\begin{align}
\cH 
\quad \rightarrow \quad 
\cO^T \cH \cO
\, ,
\qquad\qquad\qquad
\cO\in \OO(n,n)
\, .
\end{align}

\medskip 

Let us now review some illustrative examples of $T$-folds 
that have been studied in the literature.

\subsection{A toy example}

We start by reviewing a toy model example that involves several duality transformations 
of a given background. 
This example has been discussed in \cite{Kachru:2002sk,Shelton:2005cf}. 
To be pedagogical and provide simple exercises, this subsection presents geometric cases 
like a twisted torus and a torus with $H$-flux before introducing a $T$-fold example. 

\subsubsection*{Twisted torus}

Let us consider the metric of a twisted torus,
\begin{align}
 \rmd s^2 
 = 
 \rmd x^2 + \rmd y^2 + (\rmd z-m\,x\,\rmd y)^2 
 \, ,
 \qquad (m\in\mathbb{Z})
 \,.
\label{eq:twisted-torus-metric}
\end{align}
Note that this is not a supergravity solution for $m\neq 0$, 
but still is a useful example to reveal a non-geometric global property. 
As this background has isometries along $y$ and $z$ directions, 
these directions can be compactified with certain boundary conditions. For example, let us take
\begin{align}
 (x,\,y,\,z) \sim (x,\,y+1,\,z)\,,\qquad 
 (x,\,y,\,z) \sim (x,\,y,\,z+1)\,. 
\end{align}
Apparently, there is no isometry along the $x$ direction, 
but there actually exists a deformed Killing vector,
\begin{align}
 k = \partial_x + m\,y\,\partial_z \,. 
\end{align}
Thus, this isometry direction can be compactified as
\begin{align}
 (x,\,y,\,z) \sim \Exp{k}(x,\,y,\,z)=(x+1,\,y,\,z+m\,y)\,.
\label{eq:twisted-torus-identification}
\end{align}
According to this identification, a 1-form $e_z\equiv \rmd z-m\,x\,\rmd y$ 
is globally well-defined \cite{Kachru:2002sk}, 
and the metric \eqref{eq:twisted-torus-metric} is also globally well-defined. 

\medskip 

When this background is regarded as a 2-torus $T^2_{y,z}$ fibered over a base $\rmS^1_x$\,, 
the metric of the 2-torus takes the form
\begin{align}
 (\CG_{mn}) = 
 \begin{pmatrix}
 1 & -m\,x \\
 0 & 1
 \end{pmatrix}
 \begin{pmatrix}
 1 & 0 \\
 0 & 1 
 \end{pmatrix}
 \begin{pmatrix}
 1 & 0 \\
 -m\,x & 1 
 \end{pmatrix} \,. 
\end{align}
Then, as one moves around the base $\rmS_x^1$ , the metric is transformed 
by a $\GL(2)$ rotation. That is to say, for $x\to x+1$, the metric is given by
\begin{align}
 \CG_{mn}(x+1) = \bigl[\Omega^\rmT\,\CG(x)\,\Omega\bigr]_{mn} \,,\qquad
 \Omega^m{}_n \equiv
 \begin{pmatrix}
 1 & 0 \\
 -m & 1 
 \end{pmatrix} \,. 
\end{align}
This monodromy twist can be compensated by a coordinate transformation
\begin{align}
 y = y'\,, \qquad z = z'+m\,y'\,.
\end{align}
Thus the metric is single-valued up to the above coordinate transformation. 
Then this background can be understood to be geometric 
because general coordinate transformations belong to the gauge group of supergravity.

\subsubsection*{Torus with $H$-flux}

When a $T$-duality is formally performed on the twisted torus \eqref{eq:twisted-torus-metric} 
along the $x$ direction, we obtain the following background 
\begin{align}
 \rmd s^2 = \rmd x^2+\rmd y^2+\rmd z^2\,,\qquad B_2=-m\,x\,\rmd y\wedge\rmd z\,,
\end{align}
equipped with the $H$-flux,
\begin{align}
 H_3=\rmd B_2 = -m\,\rmd x\wedge\rmd y\wedge\rmd z\,.
\end{align}
If we consider the generalized metric \eqref{eq:Hmetric} on the doubled torus 
$(y,z,\tilde{y},\tilde{z})$ associated to this background, 
then we can easily identify the induced monodromy when $x \to x+1$. 
In this case, the monodromy matrix is given by
\begin{align}
 \cH_{MN}(x+1) = \bigl[\Omega^\rmT\,\cH(x)\,\Omega\bigr]_{MN}\,,\qquad 
 \Omega^M{}_N = \begin{pmatrix}
 \delta^m_n & 0 \\
 2\,m\,\delta_{[m}^y\,\delta_{n]}^z & \delta_m^n
 \end{pmatrix} \in \OO(2,2;\mathbb{Z})\,.
\end{align}
Then, the induced monodromy can be compensated by a constant shift in the $B$-field,
\begin{align}
 B_{yz} ~~\to~~ B_{yz}-m. 
\end{align}
This shift transformation, which makes the background single-valued, 
belongs to the gauge transformations of supergravity. 
Hence we conclude that the background is geometric.

\subsubsection*{$T$-fold}

Finally, let us perform another $T$-duality transformation along the $y$-direction 
on the twisted torus \eqref{eq:twisted-torus-metric}. 
Then we obtain the following background \cite{Kachru:2002sk}:
\begin{align}
 \rmd s^2 = \rmd x^2 + \frac{\rmd y^2+\rmd z^2}{1+m^2\,x^2} \,, 
\qquad B_2= \frac{m\,x}{1+m^2\,x^2}\,\rmd y\wedge\rmd z\,. 
 \label{eq:H-Q-flux}
\end{align}
In this case, neither general coordinate transformations nor $B$-field gauge transformations 
are enough to remove the multi-valuedness of the background. 
This can also be seen by calculating the monodromy matrix. 
The associated generalized metric is given by
\begin{align}
 \cH(x)=
\begin{pmatrix}
 \delta_m^p & 0 \\
 -2\,m\,x\,\delta_y^{[m}\,\delta_z^{p]} & \delta^m_p
\end{pmatrix}
\begin{pmatrix}
 \delta_{pq} & 0 \\
 0 & \delta^{pq} 
\end{pmatrix}
\begin{pmatrix}
 \delta^q_n & 2\,m\,x\,\delta_y^{[q}\,\delta_z^{n]} \\
 0 & \delta_q^n
\end{pmatrix} \,. 
\label{eq:H-Q-flux-simple}
\end{align}
Then, we find that, upon the transformation $x\to x+1$, the induced monodromy is 
\begin{align}
 \cH_{MN}(x+1) = \bigl[\Omega^\rmT\,\cH(x)\,\Omega\bigr]_{MN} \,,
\qquad \Omega^M{}_N \equiv \begin{pmatrix}
 \delta^m_n & 2\,m\,\delta_y^{[m}\,\delta_z^{n]} \\
 0 & \delta_m^n
\end{pmatrix}\in \OO(2,2;\mathbb{Z})\,. 
\label{eq:monodromy-simple}
\end{align}
The present $\OO(2,2;\mathbb{Z})$ monodromy matrix $\Omega$ takes an upper-triangular form 
(called a $\beta$-transformation) which is not part of the gauge group of supergravity. 
Hence, to keep the background globally well defined, the transition functions that glue the local patches 
should be extended to the full set of $\OO(2,2;\mathbb{Z})$ transformations beyond 
general coordinate transformations and B-field gauge transformations. 
This is what happens to the $T$-fold case. 

\medskip

In summary, we conclude that a non-geometric background 
with a non-trivial $\OO(n,n;\mathbb{Z})$ monodromy transformation, 
such as a $\beta$-transformation, is a $T$-fold. 
The background \eqref{eq:H-Q-flux} is a simple example.

\medskip 

From a viewpoint of DFT, by choosing a suitable solution of the section condition, 
the $\beta$-transformations can be realized as the gauge symmetries. 
Indeed, the above $\OO(2,2;\mathbb{Z})$ monodromy matrix $\Omega$ can be canceled 
by a generalized coordinate transformation on the double torus coordinates $(y,z,\tilde y, \tilde z)$,
\begin{align}
 y=y'+m\,\tilde{z}'\,,\qquad z=z'\,,\qquad \tilde{y}=\tilde{y}'\,,\qquad \tilde{z}=\tilde{z}' \,.
\label{eq:gen-coord-transf-simple}
\end{align}
In this sense, the twisted doubled torus is globally well-defined in DFT.

\medskip 

In addition, it is also possible to make the single-valuedness manifest by introducing the dual fields $\OG_{mn}$ 
and $\beta^{mn}$ \cite{Shapere:1988zv,Giveon:1988tt,Duff:1989tf,Tseytlin:1990nb,Giveon:1994fu} 
defined by
\begin{align}
 (\OG^{-1}+\beta)^{mn} \equiv (E^{-\rmT})^{mn} \,,\quad E_{mn}\equiv \CG_{mn}+B_{mn}\,,
\end{align}
or equivalently,
\begin{align}
 \OG_{mn} = E_{mk}\,E_{nl}\,\CG^{kl} = (\CG-B\,\CG^{-1}\,B)_{mn}\,, 
\qquad 
 \beta^{mn} = (E^{-\rmT})^{mk}\,(E^{-\rmT})^{nl}\,B_{kl} \,. 
\label{eq:dual-fields}
\end{align}
The dual metric $\OG_{mn}$ is precisely the same as the open-string metric \cite{Seiberg:1999vs}, 
and the original metric $\CG_{mn}$ may be called the closed-string metric. 
In terms of these fields, the generalized metric can be parameterized as 
(see for example \cite{Andriot:2011uh})
\begin{align}
\begin{split}
 &\cH = \Exp{\bbeta^\rmT}\check{\cH}\Exp{\bbeta} = \begin{pmatrix} \OG_{mn} & \OG_{mk}\,\beta^{kn} \\
 -\beta^{mk}\,\OG_{kn} & (\OG^{-1}-\beta\,\OG\,\beta)^{mn}
 \end{pmatrix} \,,
\\
 &\check{\cH}\equiv \begin{pmatrix}
 \OG_{mn} & 0 \\
 0 & \OG^{mn}
 \end{pmatrix}\,,\quad
 (\bbeta^M{}_N) \equiv \begin{pmatrix}
 0 & \beta^{mn} \\
 0 & 0
 \end{pmatrix}\,,\quad 
 \Exp{\bbeta} = \begin{pmatrix}
 \delta^m_n & \beta^{mn} \\
 0 & \delta_m^n
 \end{pmatrix} \,, 
\end{split}
\label{eq:non-geometric-parameterization}
\end{align}
which is referred to as a non-geometric parameterization of the generalized metric. 
At the same time, the parameterization of the DFT dilaton is also changed 
by introducing the dual dilaton $\tilde{\phi}$,
\begin{align}
 \Exp{-2d}= \Exp{-2\tilde{\phi}}\sqrt{\abs{\OG}}\,. 
\end{align}

\medskip 

In the non-geometric parameterization, the background \eqref{eq:H-Q-flux-simple} becomes
\begin{align}
 \rmd s_{\text{dual}}^2 \equiv \OG_{mn}\,\rmd x^m\,\rmd x^n 
= \rmd x^2+\rmd y^2+\rmd z^2\,,\qquad \beta^{yz} = m\,x \,,
\end{align}
and the $\OO(2,2;\mathbb{Z})$ monodromy matrix \eqref{eq:monodromy-simple} corresponds to 
a constant shift in the $\beta$ field; $\beta^{yz}\to \beta^{yz} + m$\,. 
Namely, up to a constant $\beta$-shift, which is a gauge symmetry 
\eqref{eq:gen-coord-transf-simple} of DFT, the background becomes single-valued. 

\medskip 

In this paper, we define a non-geometric $Q$-flux as \cite{Grana:2008yw}
\begin{align}
 Q_p{}^{mn} \equiv \partial_p \beta^{mn} \,.
\end{align}
Then, upon a transformation $x \to x+1$, the induced monodromy on the $\beta$-field is measured 
by an integral of the $Q$-flux,
\begin{align}
 \beta^{mn}(x+1)-\beta^{mn}(x) = \int_x^{x+1} \rmd x'^p\,\partial_p \beta^{mn}(x') 
= \int_x^{x+1} \rmd x'^p\,Q_p{}^{mn}(x') \,. 
\end{align}
This expression plays the central role in our argument. 

\medskip 

After this illustrative example we conclude that $Q$-flux backgrounds are globally well-defined 
as $T$-folds. In the next subsection, let us explain a codimension-1 example 
of the exotic $5_2^2$-brane by using the above $Q$-flux. 

\subsection{Codimension-1 $5^2_2$-brane background}

The second example is a supergravity solution studied in \cite{Hassler:2013wsa}. 
It is obtained by smearing the codimension-2 exotic $5^2_2$-brane solution 
\cite{LozanoTellechea:2000mc,deBoer:2010ud}, which is related to the NS5-brane solution 
by two $T$-duality transformations. 
It is also referred to as a $Q$-brane, as it is a source of $Q$-flux, as we are going to check. 
The codimension-1 version of this solution is given by
\begin{align}
\begin{split}
 \rmd s^2 &= m\, x\,(\rmd x^2+\rmd y^2) + \frac{x\,(\rmd z^2+\rmd w^2)}{m\,(x^2+z^2)} 
+ \rmd s_{\mathbb{R}^6}^2 \,,
\\
 B_2&= \frac{x}{m\,(x^2+z^2)}\rmd z\wedge\rmd w \,,\qquad \Phi 
= \frac{1}{2}\,\ln\biggl[\frac{x}{m\,(x^2+z^2)}\biggr] \,. 
\end{split}
\end{align}
With the non-geometric parameterization \eqref{eq:non-geometric-parameterization}, 
this solution is simplified as 
\begin{align}
\begin{split}
 \rmd s_{\text{dual}}^2 &= m\, x\,(\rmd x^2+\rmd y^2) 
+ \frac{\rmd z^2+\rmd w^2}{m\,x} + \rmd s_{\mathbb{R}^6}^2 \,, 
\\
 \beta^{zw} &=m\,y \,,\qquad \tilde{\phi} = \frac{1}{2}\,\ln \biggl[\frac{1}{m\,x}\biggr] \,. 
\end{split}
\end{align}
Assuming that the $y$ direction is compactified with $y\sim y+1$, 
the monodromy under $y\to y+1$ is given by a constant $\beta$-shift; 
\begin{align}
 \beta^{zw}\to \beta^{zw}+m\,.
\end{align}
As the background is twisted by a $\beta$-shift, this example can be considered as a $T$-fold. 
In terms of the $Q$-flux, this solution has a constant $Q$-flux,
\begin{align}
 Q_y{}^{zw} = m \,. 
\end{align}
Finally, the monodromy matrix is given by
\begin{align}
 \cH_{MN}(y+1) = \bigl[\Omega^\rmT\,\cH(y)\,\Omega\bigr]_{MN} \,,\qquad 
 \Omega^M{}_N \equiv \begin{pmatrix}
 \delta^m_n & 2\,m\,\delta_z^{[m}\,\delta_w^{n]} \\
 0 & \delta_m^n
\end{pmatrix}\in \OO(10,10;\mathbb{Z})\,. 
\end{align}

\medskip

By employing the knowledge on $T$-folds introduced in this section, 
we will elaborate on a non-geometric aspect of YB-deformed backgrounds as $T$-folds.

\section{Non-geometric aspects of YB deformations}
\label{sec:non-geometry}

Let us show that various YB-deformed backgrounds can be regarded as $T$-folds. 

\medskip 

Subsec.\,\ref{sec:review-GSE} is devoted to 
a brief review of the generalized supergravity to fix our convention and notation. 
In Subsec.\,\ref{sec:YB-beta-deform}, we explain how the homogeneous Yang-Baxter deformations 
are interpreted as $\beta$-twists 
and how a YB-deformed background can be derived from a given classical $r$-matrix. 
In Subsec.\,\ref{sec:monodromy-YB}, the general structure of $T$-duality monodromy is revealed 
for the YB-deformed backgrounds studied in this paper. In Subsec.\,\ref{sec:non-geometry-Minkowski}, 
various $T$-folds are obtained as YB-deformations of Minkowski spacetime. 
In Subsec.\,\ref{sec:non-geometry-NATD}, we study a certain background which is obtained 
by a non-Abelian $T$-duality but is not described as a Yang-Baxter deformation. It is shown that 
this background is a solution of GSE and can also be regarded as a $T$-fold. 
In Sec.\,\ref{sec:non-geometry-AdS5xS5}, in order to study a more non-trivial class of $T$-folds 
with R-R fields, we consider some backgrounds obtained as YB-deformations of $\AdS5\times\rmS^5$. 

\subsection{Generalized supergravity}
\label{sec:review-GSE}

The generalized type IIB supergravity equations of motion were originally derived 
in \cite{Arutyunov:2015mqj,Wulff:2016tju}. Just for later convenience, we will follow the convention 
utilized in \cite{Sakamoto:2017wor} hereafter. 

\medskip 

Then the generalized type II supergravity equations of motion are given by
\begin{align}\label{GSE}
 &R_{mn}- \frac{1}{4}\,H_{mpq}\,H_n{}^{pq} + 2\,D_m \partial_n \Phi + D_m U_n +D_n U_m = T_{mn} \,,
\nn\\
 &R + 4\,D^m \partial_m \Phi - 4\,\abs{\partial \Phi}^2 - \frac{1}{2}\,\abs{H_3}^2 
- 4\,\bigl(I^m I_m+U^m U_m + 2\,U^m\,\partial_m \Phi - D_m U^m\bigr) =0 \,,
\nn\\
 &-\frac{1}{2}\,D^k H_{kmn} + \partial_k\Phi\,H^k{}_{mn} + U^k\,H_{kmn} + D_m I_n - D_n I_m = \cK_{mn} \,,
\\
 &\rmd *\hat{\sfF}_p - H_3\wedge * \hat{\sfF}_{p+2} - \iota_I B_2 \wedge * \hat{\sfF}_p -\iota_I * \hat{\sfF}_{p-2} =0 \,,
\nn
\end{align}
where $I=I^m\,\partial_m$ is a Killing vector satisfying
\begin{align}
 \Lie_I \CG_{mn} = D_m I_n + D_n I_m = 0\,. 
\end{align}
Here $D_m$ is the covariant derivative associated with the metric $\CG_{mn}$, 
$*$ is the Hodge star operator, and $\iota_I$ is the interior product with the vector $I$. 
In addition, we have introduced the following quantities: 
\begin{align}
\begin{split}
 T_{mn} &\equiv \frac{1}{4}\Exp{2\Phi} \sum_p \Bigl[ \frac{1}{(p-1)!}\, 
 \hat{\sfF}_{(m}{}^{k_1\cdots k_{p-1}} \hat{\sfF}_{n) k_1\cdots k_{p-1}} - \frac{1}{2}\, 
 \CG_{mn}\,\abs{\hat{\sfF}_p}^2 \Bigr] \,,
\\
 \cK_{mn}&\equiv \frac{1}{4}\Exp{2\Phi} \sum_p \frac{1}{(p-2)!}\, \hat{\sfF}_{k_1\cdots k_{p-2}}\, 
 \hat{\sfF}_{mn}{}^{k_1\cdots k_{p-2}} \,,\qquad U_m \equiv I^n\,B_{nm} \,. 
\end{split}
\end{align}
Here, $0\leq p\leq 9$ takes an even/odd number for type IIA/IIB theory, respectively. 
The R-R field strengths should satisfy the self-duality relation,
\begin{align}
 * \hat{\sfF}_p = (-1)^{\frac{p(p+1)}{2}+1} \hat{\sfF}_{10-p} \,,\qquad 
 \hat{\sfF}_p = (-1)^{\frac{p(p-1)}{2}} * \hat{\sfF}_{10-p} \,. 
\end{align}
Given the R-R field strengths, the R-R potentials can be determined through the relation,
\begin{align}
 \hat{\sfF}_p = \rmd \hat{\sfC}_{p-1} + H_3\wedge \hat{\sfC}_{p-3} 
- \iota_I B_2 \wedge \hat{\sfC}_{p-1} -\iota_I \hat{\sfC}_{p+1} \,. 
\end{align}
Note that when $I=0$, the above expressions reduce to those of the usual supergravity. 

\medskip

It is also convenient to define the R-R fields $(\sfF,\,\sfA)$ and $(\check{\sfF},\,\check{\sfC})$ as
\begin{align}
 \sfF \equiv \Exp{B_2\wedge} \hat{\sfF} \,,\qquad \sfA \equiv \Exp{B_2\wedge} \hat{\sfC} \,, \qquad
 \check{\sfF} \equiv \Exp{\beta\vee} \sfF \,,\qquad 
 \check{\sfC} \equiv \Exp{\beta\vee} \sfA \,,
\end{align}
satisfying
\begin{align}
\begin{split}
 \sfF_p &= \rmd \sfA_{p-1} -\iota_I \sfA_{p+1} \,,
\\
 \check{\sfF}_p &= \rmd \check{\sfC}_{p-1} -\frac{1}{2}\,Q^{mn}\wedge \iota_m \iota_n 
\check{\sfC}_{p+1} -\iota_I \check{\sfC}_{p+1} \qquad \bigl(Q^{mn}\equiv Q_k{}^{mn}\,\rmd x^k\bigr)\,. 
\end{split}
\end{align}
Here, for a bi-vector $\beta^{mn}$ and a $p$-form $\alpha_p$, we have defined
\begin{align}
 \beta\vee \alpha_p \equiv \frac{1}{2}\,\beta^{mn}\,\iota_m\,\iota_n \alpha_p \,. 
\end{align}

\medskip 

In order to distinguish three definitions of R-R fields, 
we call $(\hat{\sfF},\,\hat{\sfC})$ $B$-untwisted R-R fields while $(\check{\sfF},\, 
\check{\sfC})$ $\beta$-untwisted R-R fields,
\begin{align}
 (\check{\sfF},\,\check{\sfC}) \quad \xleftarrow[\quad \beta\text{-untwist}\quad]{\Exp{\beta\vee}}\quad 
 (\sfF,\,\,\sfA) \quad \xrightarrow[\quad B\text{-untwist}\quad]{\Exp{-B_2\wedge}}\quad 
 (\hat{\sfF},\,\hat{\sfC}) \,.
\end{align}
Following the same terminology, we call the dual metric and the dual dilaton 
$(\OG_{mn},\,\tilde{\phi})$ the $\beta$-untwisted fields,
\begin{gather}
\begin{aligned}
 \cH&= \Exp{-\bB^\rmT}\hat{\cH}\Exp{-\bB} \quad \xrightarrow[\quad B\text{-untwist}\quad]{}\quad 
 \hat{\cH} = \begin{pmatrix} \CG_{mn} & 0 \\ 0 & \CG^{mn} \end{pmatrix}
\nn\\[-5mm]
 \check{\cH} = \begin{pmatrix} \OG_{mn} & 0 \\ 0 & \OG^{mn} \end{pmatrix} \quad 
\xleftarrow[\quad \beta\text{-untwist}\quad]{}\quad &= \Exp{\bbeta^\rmT}\check{\cH}\Exp{\bbeta} 
\end{aligned}
\\
 \Exp{-2\tilde{\phi}} \quad \xleftarrow[\quad \beta\text{-untwist}\quad]{}\quad 
 \sqrt{\abs{\OG}}\Exp{-2\tilde{\phi}}=\Exp{-2d} = \sqrt{\abs{\CG}}\Exp{-2\Phi} \quad 
\xrightarrow[\quad B\text{-untwist}\quad]{}\quad 
 \Exp{-2\Phi} \,.
\end{gather}
The $B$-untwisted fields are invariant under $B$-field gauge transformations 
while the $\beta$-untwisted fields are invariant under $\beta$-transformations. 

\medskip 

When $I=0$, the $B$-untwisted fields $(\hat{\sfF},\,\hat{\sfC},\,\CG_{mn},\,\Phi)$ 
together with $B_{mn}$ are frequently utilized in some contexts. 
For example, these are the background fields appearing in the Green-Schwarz superstring action. 
On the other hand, the $\beta$-untwisted fields 
$(\check{\sfF},\,\check{\sfC},\,\OG_{mn},\,\tilde{\phi})$ 
are unfamiliar quantities but play an important role in the context of YB deformation. 
To study the monodromy of $T$-folds, the objects in the middle, 
\emph{i.e.}, $(\sfF,\,\,\sfA,\,\cH_{MN},\,d)$ will play an important role, as we will discuss later. 

\medskip

Before closing this subsection, it is worth noting the divergence formula observed 
in \cite{Araujo:2017jkb,Araujo:2017jap,Sakamoto:2017cpu}. 
For the solutions of GSE obtained as YB deformations and a non-Abelian $T$-duality 
discussed in this paper, the Killing vector $I$ can always be found from the following formula: 
\begin{align}
 I^m=\tilde{D}_{n} \beta^{mn}\,,
\label{eq:div-formula}
\end{align}
where $\tilde{D}$ is associated with the $\beta$-untwisted metric $\CG_{mn}$\,. 
The general proof of this expression for the general YB deformations based on the mCYBE and 
the homogeneous CYBE will be reported in the coming paper \cite{preparation}.

\subsection{YB deformations as $\beta$-deformations}
\label{sec:YB-beta-deform}

YB deformations of type IIB string theory on AdS$_5\times$S$^5$ have been presented 
in \cite{Delduc:2013qra,Kawaguchi:2014qwa}. It used to be quite a difficult problem to read off 
the full expression of YB deformed background, because it is necessary to perform supercoset 
construction but it is really complicated and the computation becomes messy. 

\medskip

In the pioneering work \cite{Arutyunov:2015qva}, the supercoset construction was done 
for the $q$-deformed AdS$_5\times$S$^5$. Then the technique was generalized to 
the homogeneous CYBE case in \cite{Kyono:2016jqy}. After these developments, this technique 
was refined in \cite{Borsato:2016ose} based on $\kappa$-symmetry. 
In the recent paper \cite{Sakamoto:2017cpu}\footnote{A similar way has been elaborated 
also from the viewpoint of the invariance of the Page form \cite{Araujo:2017enj}.}, 
a much simpler way has been proposed. 
This is a direct formula between the fields in GSE and classical $r$-matrices 
satisfying the homogeneous CYBE and relies on the divergence formula \eqref{eq:div-formula}. 
In the following, we will give a brief review of this simple formula and explain how to use it 
by taking a simple example. 

\medskip

As we introduced in Sec.\,\ref{sec:T-fold}, the $\beta$-deformations 
(or the $\beta$-transformations) belong to a particular class of $\OO(D,D)$ transformations 
under which the $\beta$-field is shifted as
\begin{align}
 \beta^{mn}(x) \to \beta'^{mn}(x) = \beta^{mn}(x) + r^{mn}(x) \qquad (r^{mn}=-r^{nm})\,,
\end{align}
while the $\beta$-untwisted fields remain invariant,
\begin{align}
 \check{\cH}' = \check{\cH}\,,\qquad
 \tilde{\phi}' =\tilde{\phi}\,,\qquad
 \check{\sfF}' =\check{\sfF}\,,\qquad
 \check{\sfC}' =\check{\sfC}\,.
\label{eq:beta-deformed-BG-DFT}
\end{align}
Unlike the $B$-field gauge transformations, the $\beta$-deformation is not a gauge transformation, 
and in general, the $\beta$-deformed background may not satisfy 
the (generalized) supergravity equations \eqref{GSE} 
even if the original background is a solution of the supergravity (or DFT). 

\medskip

Now, let us explain a relation between the $\beta$-deformation and the YB deformation.
For this purpose, we concentrate on deformations of a background with vanishing $B$-field, 
and then the $\beta$-field in the original background also vanishes. 
A homogeneous YB deformation is specified by taking a skew-symmetric classical $r$-matrix
\begin{align}
 r=\frac{1}{2}\,r^{ij}\,T_{i}\wedge T_{j} = r^{ij}\,T_i\otimes T_j\,,
 \qquad r^{ij}=-r^{ji}\,,
\label{r-mat}
\end{align}
which satisfies the homogeneous classical Yang-Baxter equation (CYBE),
\begin{align}
 f_{l_1l_2}{}^{i}\,r^{jl_1}\,r^{kl_2}+
 f_{l_1l_2}{}^{j}\,r^{kl_1}\,r^{il_2}+
 f_{l_1l_2}{}^{k}\,r^{il_1}\,r^{jl_2}=0\,.
\label{CYBE}
\end{align}
Here $r^{ij}$ is a constant skew-symmetric matrix and $T_i$'s are the elements of the Lie algebra 
$\mathfrak{g}$ associated with the bosonic isometry group $G$, satisfying commutation relations
\begin{align}
 [T_i,\,T_j]=f_{ij}{}^{k}\,T_k \qquad (f_{ij}{}^{k}~:~\mbox{the structure constant})\,. 
\label{Lie-com}
\end{align}
An important observation made in \cite{Sakamoto:2017cpu} is that a YB-deformed background 
associated with the classical $r$-matrix \eqref{r-mat} can also be generated 
by a $\beta$-deformation,
\begin{align}
 \beta^{mn}(x) \to \beta^{mn}(x) + r^{mn}(x)\,,\qquad \frac{1}{2}\,r^{mn}(x)\,
\partial_m\wedge\partial_n \equiv \eta\,r^{ij}\,\hat{T}_i(x)\wedge \hat{T}_j(x)\,.
\end{align}
Here, a real constant $\eta$ is a deformation parameter and $\hat{T}_i$ are Killing vector fields 
on the original background satisfying the same commutation relations \eqref{Lie-com}.
Since $\beta^{mn}=0$ in the undeformed background, we obtain the following expression 
\begin{align}
 \beta^{(r) mn}(x) = r^{mn}(x) = 2\,\eta\,r^{ij}\,\hat{T}^{m}_i(x)\,\hat{T}_j^{n}(x) 
\label{YB-beta}
\end{align}
for the YB-deformed background. 

\medskip

In terms of the usual supergravity fields ($\CG_{mn}$, $B_{mn}$, $\Phi$, $\hat{\sfF}$, $\hat{\sfC}$), 
the YB-deformed background can be expressed as
\begin{align}
\begin{split}
 &\CG_{mn}^{(r)}+B_{mn}^{(r)}=\bigl[(\OG^{-1}-\beta^{(r)})\bigr]^{-1}_{mn}\,,
\\
 &\Exp{-2\Phi^{(r)}} =\Exp{-2\tilde{\phi}}\sqrt{\det[\delta_n^m
-(\OG\,\beta^{(r)}\,\OG\,\beta^{(r)})_m{}^n]}\,,
\\
 &\hat{\sfF}^{(r)}=\Exp{-B_2^{(r)}\wedge}\Exp{-\beta^{(r)}\vee}\check{\sfF}\,,
\qquad
 \hat{\sfC}^{(r)}=\Exp{-B_2^{(r)}\wedge}\Exp{-\beta^{(r)}\vee}\check{\sfC} \,,
\label{YB-formula}
\end{split}
\end{align}
where the $\beta$-untwisted fields $(\OG_{mn}=\CG_{mn},\,\tilde{\phi}
=\Phi,\,\check{\sfF}=\hat{\sfF},\,\check{\sfC}=\hat{\sfC})$ are the original undeformed background 
with $B_2=0$. The deformed background solves the (generalized) supergravity equations of motion 
\eqref{GSE}. In this way, we can generate YB-deformed backgrounds by using the formula 
\eqref{YB-formula} with the $\beta$-field \eqref{YB-beta}.

\medskip

Furthermore, it is interesting to note that the homogeneous CYBE \eqref{CYBE} can also be expressed as
\begin{align}
 R\equiv [\beta^{(r)},\,\beta^{(r)}]_S=0\,,
\label{Jacobi}
\end{align}
where $[\,,\,]_S$ denotes the Schouten bracket and the tri-vector $R$ is known 
as the non-geometric $R$-flux. 
The Schouten bracket is defined for a $p$-vector and a $q$-vector as
\begin{align}
\begin{split}
 &[a_1\wedge \cdots \wedge a_p,\,b_1\wedge \cdots \wedge b_q]_{\rmS} \\ 
\equiv &\sum_{i,j}(-1)^{i+j} [a_i,\,b_j] \wedge a_1\wedge \cdots \check{a_i}\cdots \wedge a_p 
\wedge b_1\wedge \cdots \check{b_j}\cdots \wedge b_q \,,
\end{split}
\end{align}
where the check $\check{a_i}$ denotes the omission of $a_i$\,. 
This fact implies that the non-geometric $R$-flux vanishes 
for the homogeneous YB-deformed backgrounds 
(as far as the undeformed background has vanishing $B$-field). 

\subsubsection*{Minkowski and $\AdS5\times\rmS^5$ backgrounds}

In the following subsections, we will consider YB deformations of $10$D Minkowski spacetime 
and the $\AdS5\times\rmS^5$ background.
Before presenting various examples, we will introduce the coordinate systems and 
show the explicit form of the Killing vector fields $\hat{T}_i$ 
in $10$D Minkowski spacetime and the $\AdS5\times\rmS^5$ background.

\medskip

For a $10$D Minkowski spacetime, we take the standard Minkowski metric,
\begin{align}
 \rmd s^2_{\rm Min} =\eta_{mn}\,\rmd x^m\, \rmd x^n \qquad (m,n=0,1,\dotsc,9)\,,
\end{align}
where $\eta_{mn}=\text{diag}(-1,+1,\dotsc,+1)$\,.
In this coordinate system, the Killing vector fields $\{\hat{T}_i\}=\{\hat{P}_m,\,\hat{M}_{mn}\}$ 
are expressed as
\begin{align}
 \hat{P}_{m}=-\partial_{m}\,,\qquad
 \hat{M}_{mn}=x_{m}\,\partial_{n}-x_{n}\,\partial_{m}\,.
\end{align}
These vector fields realize the following Poincar\'e algebra:
\begin{align}
\begin{split}
 [P_{m},\,M_{nk}]&=\eta_{mn}\,P_{k}-\eta_{mk}\,P_{n}\,,
\\
 [M_{mn},\,M_{kl}]&= -\eta_{mk}\,M_{nl}+\eta_{nk}\,M_{ml} 
 + \eta_{ml}\,M_{nk}-\eta_{nl}\,M_{mk}\,.
\label{Poincare-Killing}
\end{split}
\end{align}
Here $P_{m}$ and $M_{mn}$ are the translation and Lorentz generators of the Poincar\'e group 
$\text{ISO}(1,9)$\,.

\medskip

When we consider the $\AdS5\times\rmS^5$ background as the original background, 
we choose the following coordinate system:
\begin{align}
\begin{split}
 &\rmd s^2_{\AdS{5}\times \rmS^5}=\rmd s^2_{\AdS{5}}+\rmd s_{\rmS^5}^2\,,
\\
 &\rmd s^2_{\AdS{5}} =\frac{\rmd z^2-(\rmd x^0)^2+(\rmd x^1)^2
 +(\rmd x^2)^2+(\rmd x^3)^2}{z^2}\,,
\\
 &\rmd s_{\rmS^5}^2 =\rmd r^2 + \sin^2 r\, \rmd\xi^2 + \cos^2\xi\,\sin^2 r\, \rmd\phi_1^2 
+ \sin^2r\,\sin^2\xi\, \rmd\phi_2^2 + \cos^2r\, \rmd\phi_3^2\,.
\end{split}
\end{align}
The R-R $5$-form field strength in the $\AdS5\times\rmS^5$ background is given by
\begin{align}
 \hat{\sfF}_5 &= 4\,\bigl(\omega_{\AdS5}+\omega_{\rmS^5}\bigr) \,, \qquad 
 \omega_{\AdS5} = \bar{*}\, \omega_{\rmS^5}\,,
\label{unRR5}
\end{align}
where the volume forms $\omega_{\AdS5}$ and $\omega_{\rmS^5}$ are defined as, respectively, 
\begin{align}
\begin{split}
 &\omega_{\AdS5} = - \frac{\rmd z\wedge \rmd x^0\wedge\rmd x^1\wedge 
\rmd x^2\wedge\rmd x^3}{z^5}\,, \\
 &\omega_{\rmS^5} = \sin^3r \cos r \sin\xi \cos\xi\,\rmd r\wedge \rmd 
\xi\wedge\rmd \phi_1\wedge \rmd\phi_2\wedge\rmd \phi_3\,,
\end{split}
\end{align}
and $\bar{*}$ is the Hodge star operator associated 
with the undeformed $\AdS5\times\rmS^5$ background,
\begin{align}
 (\bar{*} \alpha_p)_{n_1\cdots n_2n_3n_4n_{10-p}} 
=\frac{1}{p!}\,\varepsilon^{m_1\cdots m_p}{}_{n_1\cdots n_2n_3n_4n_{10-p}}\,\alpha_{m_1\cdots m_p}\,,
\qquad \varepsilon_{z0123r\xi\phi_1\phi_2\phi_3} = + \sqrt{\abs{\CG}}\,.
\label{eq:epsilon}
\end{align}
It is also convenient to define $\omega_4$ as
\begin{align}
 \omega_4 \equiv \sin^4r\sin\xi\cos\xi\, \rmd \xi\wedge\rmd \phi_1\wedge 
\rmd\phi_2\wedge\rmd \phi_3\,. 
\end{align}
Note that $\rmd \omega_4=4\,\omega_{\rmS^5}$\,. 

\medskip

The non-vanishing commutation relations for the isometry group $\text{SO}(2,4)$ of $\AdS5$ are
\begin{align}
\begin{split}
 [P_{\mu},\,K_{\nu}]&=2\,(M_{\mu\nu}+\eta_{\mu\nu}\, D)\,,\quad
 [D,\,P_\mu]=P_{\mu}\,,\quad
 [D,\,K_{\mu}]=-K_{\mu}\,,
\\
 [P_{\mu},\,M_{\nu\rho}]&=\eta_{\mu\nu}\,P_{\rho}-\eta_{\mu\rho}\,P_{\nu}\,,\quad
 [K_{\mu},\,M_{\nu\rho}]=\eta_{\mu\nu}\,K_{\rho}-\eta_{\mu\rho}\,K_{\nu}\,,
\\
 [M_{\mu\nu},\,M_{\rho\sigma}]&= \eta_{\mu\sigma}\,M_{\nu\rho}+\eta_{\nu\rho}\,M_{\mu\sigma}
 -\eta_{\mu\rho}\,M_{\nu\sigma}-\eta_{\nu\sigma}\,M_{\mu\rho}\,,
\label{conf-comm}
\end{split}
\end{align}
where $\mu\,,\nu=0,1,2,3$, and $D$, and $K_{\mu}$ are generators of the dilatation and special conformal transformations, respectively.
The commutation relations \eqref{conf-comm} are realized by the following Killing vector fields, $\{\hat{T}_i\}=\{\hat{P}_m,\,\hat{K}_m,\,\hat{M}_{mn},\,\hat{D}\}$:
\begin{align}
\begin{split}
 &\hat{P}_{\mu}=-\partial_{\mu}\,,\qquad
 \hat{K}_{\mu}=-(z^2+x_{\nu}\,x^{\nu})\,\partial_\mu+2\,x_{\mu}\,(z\,\partial_z+x^{\nu}\,\partial_{\nu})\,,
\\
 &\hat{M}_{\mu\nu}=x_{\mu}\,\partial_{\nu}-x_{\nu}\,\partial_{\mu}\,,\qquad 
 \hat{D}=-z\,\partial_z -x^{\mu}\,\partial_{\mu}\,.
\label{conformal-Killing}
\end{split}
\end{align}

\medskip

We will use the above Killing vector fields \eqref{Poincare-Killing}, \eqref{conformal-Killing} 
to obtain an explicit expression of the $\beta$-field $\beta^{(r)}=\eta\,r^{ij}\,
\hat{T}_i\wedge \hat{T}_j$ associated with a given $r$-matrix 
$r=\frac{1}{2}\,r^{ij}\,T_i\wedge T_j$. 
In the following, we will omit the superscript ${}^{(r)}$ 
for the YB-deformed backgrounds. 

\subsubsection*{An example: the Maldacena-Russo background}

To demonstrate how to use the formula \eqref{YB-formula},
let us consider a YB-deformed $\AdS{5}\times S^5$ background associated 
with a classical $r$-matrix \cite{Matsumoto:2014gwa},
\begin{align}
 r=\frac{1}{2}\,P_1\wedge P_2\,. \label{MR-r}
\end{align}
This $r$-matrix is Abelian and satisfies the homogeneous CYBE \eqref{CYBE}. 
The associated YB deformed background is derived in \cite{Matsumoto:2014gwa,Kyono:2016jqy}. 

\medskip

The classical $r$-matrix \eqref{MR-r} leads to the associated $\beta$-field,
\begin{align}
 \beta =\eta\,\hat{P}_1\wedge \hat{P}_2=\eta\,\partial_1\wedge \partial_2\,.
\end{align}
Then, the $\AdS{5}$ part of a $10\times 10$ matrix $(G^{-1}-\beta)$ is
\begin{align}
 \bigl(G^{-1}-\beta\bigr)^{mn}=
 \begin{pmatrix}
 z^2&0&0&0&0\\
 0&-z^2&0&0&0\\
 0&0&z^2&-\eta&0\\
 0&0&\eta&z^2&0\\
 0&0&0&0&z^2
 \end{pmatrix} \,,
\label{G-beta}
\end{align}
where we have ordered the coordinates as $(z\,,x^0\,,x^1\,,x^2\,,x^3)$\,.
By using the inverse of the matrix \eqref{G-beta} and the formula \eqref{YB-formula},
we obtain the NS-NS fields of the YB-deformed background,
\begin{align}
\begin{split}
 \rmd s^2&=\frac{\rmd z^2-(\rmd x^0)^2+(\rmd x^3)^2}{z^2}+\frac{z^2\,[(\rmd x^1)^2+(\rmd x^2)^2]}{z^4+\eta^2} +\rmd s_{\rm S^5}^2\,,
\\
 B_2&=\frac{\eta}{z^4+\eta^2}\,\rmd x^1\wedge \rmd x^2\,,\qquad
 \Phi=\frac{1}{2}\,\ln\biggl[\frac{z^4}{z^4+\eta^2}\biggr]\,.
\label{MR-NS}
\end{split}
\end{align}

\medskip

The next task is to derive the R-R fields of the deformed background.
From the undeformed R-R $5$-form field strength \eqref{unRR5} of the $\AdS5\times\rmS^5$ background,
the R-R fields $\sfF$ are given by
\begin{align}
\begin{split}
 \sfF&=\Exp{-\beta\vee}\check{\sfF}^{(0)}
 =4\,\bigl(\omega_{\AdS5}+\omega_{\rmS^5}\bigr)-4\,\beta\vee\omega_{\AdS5}
\\
 &=4\,\bigl(\omega_{\AdS5}+\omega_{\rmS^5}\bigr)-4\,\eta\,\frac{\rmd z\wedge \rmd x^0\wedge \rmd x^3}{z^5}\,.
\end{split}
\end{align}
This is nothing but a linear combination of the deformed R-R field strengths with different rank. 
Hence we can readily read off the following expressions: 
\begin{align}
 \sfF_3=-4\,\eta\,\frac{\rmd z\wedge \rmd x^0\wedge \rmd x^3}{z^5}\,,\qquad
 \sfF_5=4\,\bigl(\omega_{\AdS5}+\omega_{\rmS^5}\bigr)\,.
\end{align}
Furthermore, the $B$-untwisted R-R fields $\hat{\sfF}$ can be computed as
\begin{align}
 \hat{\sfF}&=\Exp{-B_2\wedge}\sfF
\no\\
 &=-4\,\eta\,\frac{\rmd z \wedge \rmd x^0\wedge \rmd x^3}{z^5}
 +4\,\biggl(\frac{z^4}{z^4+\eta^2}\,\omega_{\AdS5}+\omega_{\rmS^5}\biggr)
 -4\,B_2\wedge\omega_{\rmS^5}\,.
\end{align}
Namely, we obtain
\begin{align}
\begin{split}
 \hat{\sfF}_1&=0\,,\qquad \hat{\sfF}_3=-4\,\eta\,\frac{\rmd z \wedge \rmd x^0\wedge \rmd x^3}{z^5}\,,
\\
 \hat{\sfF}_5&=4\,\biggl(\frac{z^4}{z^4+\eta^2}\,\omega_{\AdS5}+\omega_{\rmS^5}\biggr)\,,
\\
 \hat{\sfF}_7&=-4\,B_2\wedge\omega_{\rmS^5}\,.
\label{MR-RR}
\end{split}
\end{align}
The full deformed background, given by \eqref{MR-NS} and \eqref{MR-RR}, is a solution of 
the standard type IIB supergravity. This background is nothing but 
a gravity dual of non-commutative gauge theory \cite{Hashimoto:1999ut,Maldacena:1999mh}. 

\medskip 

Thus, nowadays, we do not have to perform supercoset construction to obtain 
the full expression of YB-deformed background. 
Just by using a simple formula \eqref{YB-formula}, given a classical $r$-matrix, 
the full background can easily be derived.

\subsection{$T$-duality monodromy of YB-deformed background}
\label{sec:monodromy-YB}

As we explained in the previous subsection, the YB-deformed background described by 
$(\cH,\,d,\,\sfF)$ always has the following structure:
\begin{align}
\begin{split}
 \cH &= \Exp{\bm{r}^\rmT}\check{\cH}^{(0)}\Exp{\bm{r}}\,,\qquad 
 d = d^{(0)} \,, \qquad 
 \sfF = \Exp{-r\vee} \check{\sfF}^{(0)} \,,
\\
 \bm{r} &\equiv \Exp{\bbeta} = \begin{pmatrix}
 \delta^m_n & r^{mn} \\
 0 & \delta_m^n
 \end{pmatrix} \,,\qquad 
 r^{mn} \equiv 2\,\eta\,r^{ij}\, \hat{T}_i^{m}\,\hat{T}_j^{n} \,,
\end{split}
\end{align}
where $(\check{\cH}^{(0)},\,d^{(0)},\,\check{\sfF}^{(0)})$ represent the undeformed background. 
In the following examples, $B$-field vanishes in the undeformed background, 
and the $\beta$-field in the YB-deformed background is given by
\begin{align}
 \beta = \frac{1}{2}\,r^{mn}\,\partial_m\wedge\partial_n = \eta\,r^{ij}\, \hat{T}_i\wedge \hat{T}_j \,.
\end{align}
At this stage, we know only the local property of the YB-deformed background. 

\medskip

In the examples considered in this paper, the bi-vector $r^{mn}$ (or the $\beta$-field in the YB-deformed background) always has a linear-coordinate dependence. 
Suppose that $r^{mn}$ depends on a coordinate $y$ linearly like,
\begin{align}
 r^{mn}=\mathsf{r}^{mn}\,y + \mathsf{\bar{r}}^{mn} \qquad (\mathsf{r}^{mn}:\text{ constant},\quad \mathsf{\bar{r}}^{mn}:\text{ independent of $y$}) \,,
\end{align}
and the $\beta$-untwisted fields are independent of $y$. 
Then, from the Abelian property,
\begin{align}
 \Exp{\bm{r_1+r_2}}=\Exp{\bm{r_1}}\Exp{\bm{r_2}}=\Exp{\bm{r_2}}\Exp{\bm{r_1}}\,,\qquad
 \Exp{-(r_1+r_2)\vee}=\Exp{-r_1\vee}\Exp{-r_2\vee}=\Exp{-r_2\vee}\Exp{-r_1\vee}\,,
\end{align}
we obtain
\begin{align}
\begin{split}
 \cH_{MN}(y+a) &= \bigl[\Omega_a^\rmT\cH(y)\,\Omega_a\bigr]_{MN} \,,\quad 
 d(y+a) = d(y) \,, \quad 
 \sfF(y+a) = \Exp{-\omega_a\vee} \check{\sfF}(y) \,,
\\
 (\Omega_a)^M{}_N &\equiv\begin{pmatrix}
 \delta^m_n & a\,\mathsf{r}^{mn} \\
 0 & \delta_m^n
 \end{pmatrix}\,,\qquad 
 (\omega_a)^{mn} \equiv a\,\mathsf{r}^{mn} \,.
\end{split}
\end{align}
If we can find out an $a_0$ (where the $\OO(10,10;\,\mathbb{R})$ matrix $\Omega_{a_0}$ 
is an element of $\OO(10,10;\mathbb{Z})$), 
the background allows us to compactify the $y$ direction as $y\sim y + a_0$\,. 
This is because $\OO(10,10;\mathbb{Z})$ is a gauge symmetry of String Theory and the background can be identified up to the gauge transformation. 
In this example of $T$-fold, the monodromy matrices for the generalized metric and R-R fields 
are $\Omega_{a_0}$ and $\Exp{-\omega_{a_0}\vee}$, respectively, 
while the dilaton $d$ is single-valued. 
Note that the R-R potential $\sfA$ has the same monodromy as $\sfF$\,.

\subsection{YB-deformed Minkowski backgrounds}
\label{sec:non-geometry-Minkowski}

In this subsection, we study YB-deformations of Minkowski spacetime 
\cite{Matsumoto:2015ypa,Borowiec:2015wua}. 
We begin by a simple example of the Abelian YB deformation. 
Then two purely NS-NS solutions of GSE are presented and are shown to be $T$-folds. 
These backgrounds have vanishing R-R fields and are the first examples of 
purely NS-NS solutions of GSE. 

\subsubsection{Abelian example}

Let us consider a simple Abelian $r$-matrix \cite{Matsumoto:2015ypa} 
\begin{align}
 r=-\frac{1}{2}\,P_1\wedge M_{23}\,. \label{Melvin-r}
\end{align}
The corresponding YB-deformed background becomes 
\begin{align}
 \rmd s^2&= -(\rmd x^0)^2+\frac{(\rmd x^1)^2+\bigl[1+(\eta\, x^2)^2\bigr]\,(\rmd x^2)^2 + \bigl[1+(\eta\, x^3)^2\bigr]\,(\rmd x^3)^2 + 2\,\eta^2\,x^2\,x^3\,\rmd x^2\,\rmd x^3}{1+\eta^2 \, \bigl[(x^2)^2+(x^3)^2\bigr]} 
\nn\\
 & \qquad +\sum_{i=4}^9(\rmd x^i)^2\,,
\nn\\
 B_2 &= \frac{\eta\,\rmd x^1\wedge \bigl(x^2\,\rmd x^3 - x^3\, \rmd x^2\bigr)}{1+\eta^2\, \bigl[(x^2)^2+(x^3)^2\bigr]}\,,\qquad
 \Phi = \frac{1}{2} \ln\biggl[\frac{1}{1+\eta^2\, \bigl[(x^2)^2+(x^3)^2\bigr]}\biggr]\,. 
 \label{Melvin}
\end{align}
It seems very messy, but after moving to an appropriate polar coordinate system 
(see Sec.\,3.1 of \cite{Matsumoto:2015ypa}), this background \eqref{Melvin} 
is found to be the well-known Melvin background \cite{Tseytlin:1994ei,Gibbons:1987ps,Hashimoto:2004pb}. 
In \cite{Matsumoto:2015ypa}, it was reproduced as a Yang-Baxter deformation 
with the classical $r$-matrix \eqref{Melvin-r}. 
For later convenience, we will keep the expression in \eqref{Melvin}. 

\medskip

The dual parameterization of this background is given by 
\begin{align}
 \rmd s^2_{\text{dual}} = -(\rmd x^0)^2 + \sum_{i=1}^9(\rmd x^i)^2 \,, \quad
 \beta = \eta\,\bigl(x^2 \,\partial_1\wedge \partial_3 -x^3\,\partial_1\wedge \partial_2 \bigr)\,, \quad 
 \tilde{\phi}=0 \,. 
\end{align}
Hence, under a shift $x^2\to x^2+\eta^{-1}$\,, the background receives the $\beta$-transformation,
\begin{align}
 \beta ~~\to~~ \beta + \partial_1\wedge \partial_3 \,. 
\end{align}
Therefore, if the $x^2$ direction is compactified with the period $\eta^{-1}$, 
then the monodromy matrix becomes
\begin{align}
 \cH_{MN}(x^2+\eta^{-1}) = \bigl[\Omega^\rmT\cH(x^2)\,\Omega\bigr]_{MN} \,,\qquad 
 \Omega^M{}_N \equiv \begin{pmatrix}
 \delta^m_n & 2\,\delta_1^{[m}\,\delta_3^{n]} \\
 0 & \delta_m^n \end{pmatrix} \in \OO(10,10;\mathbb{Z})\,. 
\end{align}
Thus this background has been shown to be a $T$-fold. 

\medskip

When the $x^3$ direction is also identified with the period $\eta^{-1}$, 
the corresponding monodromy matrix becomes
\begin{align}
 \cH_{MN}(x^3+\eta^{-1}) = \bigl[\Omega^\rmT\cH(x^3)\,\Omega\bigr]_{MN} \,,\qquad 
 \Omega^M{}_N \equiv \begin{pmatrix}
 \delta^m_n & - 2\,\delta_1^{[m}\,\delta_2^{n]} \\
 0 & \delta_m^n \end{pmatrix} \in \OO(10,10;\mathbb{Z})\,. 
\end{align}

\medskip 

In terms of non-geometric fluxes, this background has a constant $Q$-flux. 
In the examples of $T$-folds presented in Sec.\,\ref{sec:T-fold}, 
a background with a constant $Q$-flux, $Q_p{}^{mn}$, 
is mapped to another background with a constant $H$-flux, $H_{pmn}$, 
under a double $T$-duality along $x^m$ and $x^n$ directions. 
On the other hand, in the present example, the background has two types of constant $Q$-fluxes, $Q_2{}^{13}$ and $Q_3{}^{12}$\,, 
but we cannot perform $T$-dualities to make the background a constant-$H$-flux background 
because $x^2$ and $x^3$ directions are not isometry directions. 

\subsubsection{Non-unimodular example 1: \ $r = \frac{1}{2} \, (P_0-P_1) \wedge M_{01}$}
\label{sec:Minkowski-example1}

Let us consider a non-unimodular classical $r$-matrix\footnote{As far as we know, 
this example has not been discussed anywhere so far. } 
\begin{align}
 r = \frac{1}{2} \, (P_0-P_1) \wedge M_{01}\,.
\end{align}
The corresponding YB-deformed background becomes
\begin{align}
\begin{split}
 \rmd s^2 &= \frac{-(\rmd x^0)^2+(\rmd x^1)^2}{1- \eta^2\,(x^0+x^1)^2} + \sum_{i=2}^9(\rmd x^i)^2\,, 
\\ 
 B_2 &= -\frac{\eta\, (x^0+x^1)}{1-\eta^2\,(x^0+x^1)^2}\, \rmd x^0 \wedge \rmd x^1\,, \quad 
 \Phi = \frac{1}{2}\ln \biggl[\frac{1}{1-\eta^2\,(x^0+x^1)^2}\biggr]\,. 
\end{split}
\label{flatpen}
\end{align}
Apparently, this background has a coordinate singularity at $x^0+x^1=\pm 1/\eta$. 
But when the dual parameterization \eqref{eq:non-geometric-parameterization} is employed, 
the dual fields are given by 
\begin{align}
 \rmd s^2_{\text{dual}} = -(\rmd x^0)^2+\sum_{i=1}^9(\rmd x^i)^2 \,, \quad 
 \beta = \eta\,(x^0+x^1)\,\partial_0\wedge \partial_1 \,,\quad \tilde{\phi} = 0 \,,
\end{align}
and they are regular everywhere.%
\footnote{A similar resolution of singularities in the dual parameterization has been argued in \cite{Malek:2012pw,Malek:2013sp} in the context of the exceptional field theory.}

\medskip

By introducing a Killing vector $I$ with the help of the divergence formula \eqref{eq:div-formula} as 
\begin{align}
 I = \tilde{D}_{n} \beta^{mn}\,\partial_m = \partial_{n} \beta^{mn}\,\partial_m 
 = \eta\,(\partial_0 -\partial_1) \,,
\end{align}
the background \eqref{flatpen} with this $I$ solves GSE. 

\medskip

Since the $\beta$-field depends on $x^1$ linearly, as one moves along the $x^1$ direction, 
the background is twisted by the $\beta$-transformation. 
In particular, when the $x^1$ direction is identified with period $1/\eta$, 
this background becomes a $T$-fold with an $\OO(10,10;\mathbb{Z})$ monodromy,
\begin{align}
 \cH_{MN}(x^1+\eta^{-1}) = \bigl[\Omega^\rmT\cH(x^1)\,\Omega\bigr]_{MN} \,,\qquad 
 \Omega^M{}_N\equiv \begin{pmatrix} \delta^m_n & 2\,\delta_0^{[m}\,\delta_1^{n]} \\ 
 0 & \delta_m^n \end{pmatrix} \,. 
\end{align}

\medskip

Note that an arbitrary solution of GSE can be regarded as a solution of DFT \cite{Sakamoto:2017wor}. 
Indeed, by introducing the light-cone coordinates and a rescaled deformation parameter as
\begin{align}
 x^\pm \equiv \frac{x^0 \pm x^1}{\sqrt{2}} \,,\qquad \bar{\eta} = \sqrt{2}\,\eta \,,
\end{align}
the present YB-deformed background can be regarded as the following solution of DFT:
\begin{align}
 \cH = \begin{pmatrix}
 0 & -1 & -\bar{\eta}\,x^+ & 0 \\
 -1 & 0 & 0 & \bar{\eta}\, x^+ \\
 -\bar{\eta}\, x^+ & 0 & 0 & (\bar{\eta}\,x^+)^2 -1 \\
 0 & \bar{\eta}\,x^+ & (\bar{\eta}\,x^+)^2 - 1 & 0 
\end{pmatrix}\,,\qquad d=\bar{\eta}\,\tilde{x}_-\,, 
\end{align}
where only $(x^+,x^-,\tilde{x}_+,\tilde{x}_-)$-components of $\cH_{MN}$ are displayed. 
Note here that the dilaton has an explicit dual-coordinate dependence 
because we are now considering a non-standard solution of the section condition 
which makes this background a solution of GSE rather than the usual supergravity. 

\medskip

Before perfoming this YB deformation (\emph{i.e.}~$\bar{\eta}=0$), 
there is a Killing vector $\chi \equiv\partial_+$\,, 
but the associated isometry is broken for non-zero $\bar{\eta}$\,. 
However, even after deforming the geometry, there exists a generalized Killing vector
\begin{align}
 \chi \equiv \Exp{\bar{\eta}\,\tilde{x}_-}\partial_+ \qquad 
 \bigl(\gLie_{\chi}\cH_{MN}=0\,,\quad \gLie_{\chi}d=0\bigr) \,,
\end{align}
which goes back to the original Killing vector in the undeformed limit, $\bar{\eta}\to 0$\,. 
In order to make the generalized isometry manifest, 
let us consider a generalized coordinate transformation,
\begin{align}
 x'^+ = \Exp{-\bar{\eta} \,\tilde{x}_-} x^+ \,,\qquad \tilde{x}'_- 
 = -\bar{\eta}^{-1}\Exp{-\bar{\eta}\,\tilde{x}_-}\,,\qquad 
 x'^M = x^M\quad (\text{others})\,. 
\end{align}
By employing Hohm and Zwiebach's finite transformation matrix \cite{Hohm:2012gk},
\begin{align}
 \cF_M{}^N \equiv \frac{1}{2}\,\Bigl(\frac{\partial x^K}{\partial x'^M}\frac{\partial x'_K}{\partial x_N}
        +\frac{\partial x'_M}{\partial x_{K}}\frac{\partial x^N}{\partial x'^K}\Bigr) \,,
\end{align}
the generalized Killing vector in the primed coordinates becomes constant, $\chi = \partial'_+$\,. 
We can also check that the generalized metric in the primed coordinate system 
is precisely the undeformed background. Namely, at least locally, 
the YB deformation can be undone by the generalized coordinate 
transformation\footnote{In the study of YB deformations of AdS$_5$\,, 
the similar phenomenon has already been observed in \cite{Orlando:2016qqu}.}. 
This fact is consistent with the fact that YB deformations can be realized 
as the generalized diffeomorphism \cite{Sakamoto:2017cpu}. 

\paragraph*{Non-Riemannian background:} 
Since the above background has a linear coordinate dependence on $\tilde{x}_-$\,, 
let us rotate the solution to the canonical section 
(\emph{i.e.}~a section in which all of the fields are independent of the dual coordinates). 
By performing a $T$-duality along the $x^-$ direction, we obtain
\begin{align}
 \cH = \begin{pmatrix}
 0 & 0 & -\bar{\eta}\, x^+ & -1 \\
 0 & 0 & (\bar{\eta}\,x^+)^2-1 & \bar{\eta}\,x^+ \\
 -\bar{\eta}\,x^+ & (\bar{\eta}\,x^+)^2 -1 & 0 & 0 \\
 -1 & \bar{\eta}\,x^+ & 0 & 0 
\end{pmatrix}\,,\qquad 
 d=\bar{\eta}\,x^- \,.
\label{eq:non-Riemannian}
\end{align}
The resulting background is indeed a solution of DFT defined on the canonical section. 
However, this solution cannot be parameterized in terms of $(\CG_{mn},\,B_{mn})$ and 
is called a non-Riemannian background in the terminology of \cite{Lee:2013hma}. 
This background does not even allow the dual parameterization \eqref{eq:non-geometric-parameterization} in terms of $(\OG_{mn},\,\beta^{mn})$\footnote{For another example of non-Riemannian backgrounds, see \cite{Lee:2013hma}. A classification of non-Riemannian backgrounds in DFT has been made in \cite{Morand:2017fnv}. In the context of the exceptional field theory, non-Riemannian backgrounds have been found in \cite{Malek:2013sp} even before \cite{Lee:2013hma}. There, the type IV generalized metrics do not allow both the conventional and dual parameterizations similar to our solution \eqref{eq:non-Riemannian}.}. 

\subsubsection{Non-unimodular example 2: \ $r=\frac{1}{2\sqrt{2}}\, \sum_{\mu=0}^4 \bigl(M_{0\mu}-M_{1\mu}\bigr) \wedge P^\mu$}
\label{sec:Minkowski-example2}

The next example is the classical $r$-matrix\cite{Borowiec:2015wua}, 
\begin{align}
 r=\frac{1}{2\sqrt{2}}\, \sum_{\mu=0}^4 \bigl(M_{0\mu}-M_{1\mu}\bigr) \wedge P^\mu\,. 
\end{align}
This classical $r$-matrix is a higher dimensional generalization of the light-cone 
$\kappa$-Poincar\'e $r$-matrix in the four dimensional one. 

\medskip

By using the light-cone coordinates,
\begin{align}
 x^\pm \equiv \frac{x^0 \pm x^1}{\sqrt{2}} \,,
\end{align}
the corresponding YB-deformed background becomes
\begin{align}
\begin{split}
 \rmd s^2&= \frac{-2\,\rmd x^+\, \rmd x^- - \eta^2\,\rmd x^+ \bigl[\sum_{i=2}^4(x^i)^2 \, \rmd x^+ - 2\,x^+ \sum_{i=2}^4x^i\,\rmd x^i \bigr]}{1-(\eta \, x^+)^2} + \sum_{i=2}^9(\rmd x^i)^2\,,
\\
 B_2&= \frac{\eta\,\rmd x^+\wedge \bigl(x^+\,\rmd x^- -\sum_{i=2}^4x^i\,\rmd x^i \bigr)}{1-(\eta \, x^+)^2} \,,\quad 
 \Phi = \frac{1}{2}\ln\biggl[\frac{1}{1-(\eta \, x^+)^2}\biggr] \,.
\end{split}
\label{eq:penHvT}
\end{align}
In terms of the dual parameterization, this background becomes
\begin{align}
\begin{split}
 &\rmd s^2_{\text{dual}} = -2\,\rmd x^+\,\rmd x^- + \sum_{i=2}^9(\rmd x^i)^2 \,,\quad \tilde{\phi}=0 \,, 
\\
 &\beta = \eta\,\sum_{\mu=0}^4 \hat{M}_{-\mu} \wedge \hat{P}^\mu = \eta\, \partial_-\wedge \bigl(x^+\,\partial_+ + {\textstyle\sum}_{i=2}^4\, x^i\, \partial_i \bigr) \,.
\end{split}
\end{align}
Again, by introducing a Killing vector from the divergence formula \eqref{eq:div-formula} as 
\begin{align}
 I = 4\,\eta\,\partial_- \,, 
\end{align}
the background \eqref{eq:penHvT} with this $I$ solves GSE. 

\medskip

This background can also be regarded as the following solution of DFT:
\begin{align}
\begin{split}
 & \cH 
 = {\tiny\begin{pmatrix}
 0 & -1 & 0 & 0 & 0 & -\eta\,x^+ & 0 & -\eta\,x^2 & -\eta\,x^3 & -\eta\,x^4 \\
 -1 & 0 & 0 & 0 & 0 & 0 & \eta\,x^+ & 0 & 0 & 0 \\
 0 & 0 & 1 & 0 & 0 & 0 & -\eta\,x^2 & 0 & 0 & 0 \\
 0 & 0 & 0 & 1 & 0 & 0 & -\eta\,x^3 & 0 & 0 & 0 \\
 0 & 0 & 0 & 0 & 1 & 0 & -\eta\,x^4 & 0 & 0 & 0 \\
 -\eta\,x^+ & 0 & 0 & 0 & 0 & 0 & (\eta\,x^+)^2 -1 & 0 & 0 & 0 \\
 0 & \eta\,x^+ & -\eta\,x^2 & -\eta\,x^3 & -\eta\,x^4 & (\eta\,x^+)^2 -1 
 & \eta^2\,\sum_{i=2}^4(x^i)^2 & \eta^2\, x^+\,x^2 & \eta^2 \, x^+\,x^3 & \eta^2\,x^+\,x^4 \\
 -\eta\,x^2 & 0 & 0 & 0 & 0 & 0 & \eta^2\, x^+\,x^2 & 1 & 0 & 0 \\
 -\eta\,x^3 & 0 & 0 & 0 & 0 & 0 & \eta^2\, x^+\,x^3 & 0 & 1 & 0 \\
 -\eta\,x^4 & 0 & 0 & 0 & 0 & 0 & \eta^2\, x^+\,x^4 & 0 & 0 & 1 
\end{pmatrix}} \,,
\\
 &d= 4\,\eta\,\tilde{x}_- \,, 
\end{split}
\end{align}
where only $(x^+,\,x^-,\,x^2,\,x^3,\,x^4,\,\tilde{x}_+,\,\tilde{x}_-,\,\tilde{x}_2,\,
\tilde{x}_3,\,\tilde{x}_4)$-components of $\cH_{MN}$ are displayed. 

\medskip

When one of the $(x^2,\,x^3,\,x^4)$-coordinates, say $x^2$, is compactified 
with the period $x^2 \sim x^2 + \eta^{-1}$, 
the monodromy matrix is given by 
\begin{align}
 \cH_{MN}(x^2 +\eta^{-1}) = \bigl[\Omega^\rmT\cH(x^2)\,\Omega\bigr]_{MN} \,,\qquad 
 \Omega^M{}_N\equiv \begin{pmatrix} \delta^m_n & 2\,\delta_-^{[m}\,\delta_2^{n]} \\ 0 & \delta_m^n \end{pmatrix} \in \OO(10,10;\mathbb{Z})\,,
\end{align}
and in this sense the compactified background is a $T$-fold. 
In terms of the non-geometric $Q$-flux, 
this background has the following components of it: 
\begin{align}
 Q_+{}^{-+} = Q_2{}^{-2} = Q_3{}^{-3} = Q_4{}^{-4} = \eta \,.
\end{align}

\subsection{A non-geometric background from non-Abelian $T$-duality}
\label{sec:non-geometry-NATD}

Before considering YB-deformations of $\AdS5\times\rmS^5$\,, let us consider another example 
of purely NS-NS background, 
which was found in \cite{Gasperini:1993nz} via a non-Abelian $T$-duality. 

\medskip 

The background takes the form,
\begin{align}
\begin{split}
 \rmd s^2&= -\rmd t^2 + \frac{(t^4+y^2)\,\rmd x^2-2\,x\,y\,\rmd x\,\rmd y+(t^4+x^2)\,\rmd y^2+t^4\,\rmd z^2}{t^2\,(t^4+x^2+y^2)} + \rmd s_{T^6}^2\,,
\\
 B_2 &= \frac{(x\,\rmd x+y\,\rmd y)\wedge \rmd z}{t^4+x^2+y^2}\,,\qquad 
 \Phi= \frac{1}{2} \ln\biggl[\frac{1}{t^2\,(t^4+x^2+y^2)}\biggr] \,, 
\end{split}
\label{eq:bgNATD}
\end{align}
where $\rmd s_{T^6}^2$ is the flat metric on a 6-torus. 
In terms of the dual parameterization, this background takes 
a Friedmann-Robertson-Walker-type form,
\begin{align}
\begin{split}
 \rmd s^2_{\text{dual}}&= -\rmd t^2 + t^{-2}\,\bigl(\rmd x^2+ \rmd y^2 
 + \rmd z^2\bigr) + \rmd s_{T^6}^2\,,
\\
 \beta &= (x\,\partial_x+y\,\partial_y)\wedge \partial_z\,,\qquad 
 \tilde{\phi} = -\ln t^3 \,. \label{3.69}
\end{split}
\end{align}
Note here that this background cannot be represented by a coset or a Lie group itself. 
This is because the background \eqref{eq:bgNATD} contains a curvature singularity 
and is not homogeneous. 
Hence the background \eqref{eq:bgNATD} cannot be realized as a Yang-Baxter deformation 
and is not included in the discussion of \cite{Hoare:2016wsk,Borsato:2016pas,Hoare:2016wca}. 

\medskip

It is easy to see that the associated $Q$-flux is constant on this background \eqref{3.69},
\begin{align}
 Q_y{}^{xy} = Q_z{}^{xz} = -1\,. 
\end{align}
Therefore, if the $x$-direction is compactified as $x\sim x+1$, 
the background fields are twisted by an $\OO(10,10;\mathbb{Z})$ transformation as
\begin{align}
 \cH_{MN}(x+1) = \bigl[\Omega^\rmT\cH(x)\,\Omega\bigr]_{MN} \,,\qquad 
 \Omega^M{}_N \equiv \begin{pmatrix} \delta^m_n & 2\,\delta_x^{[m} \,\delta_z^{n]} \\ 0 
 & \delta_m^n \end{pmatrix}\,,\quad 
 d(x+1)=d(x) \,. 
\end{align}
Thus the background can be interpreted as a $T$-fold. 
If the $z$-direction is also compactified as $z\sim z+1$, another twist is realized as 
\begin{align}
 \cH_{MN}(y+1) = \bigl[\Omega^\rmT\cH(y)\,\Omega\bigr]_{MN} \,,\qquad 
 \Omega^M{}_N \equiv \begin{pmatrix} \delta^m_n & 2\,\delta_y^{[m} \,\delta_z^{n]} \\ 
 0 & \delta_m^n \end{pmatrix}\,,\quad 
 d(y+1)=d(y) \,. 
\end{align}

\medskip

As stated in \cite{Gasperini:1993nz}, this background is not a solution of the usual supergravity. 
However, by using the divergence formula $I^m = \tilde{D}_n\beta^{mn}$ again and introducing 
a vector field as 
\begin{align}
 I=-2\,\partial_z \,,
\end{align}
we can see that the background \eqref{eq:bgNATD} together with this vector field $I$ 
satisfies GSE. 
Thus, this background can also be regarded as a $T$-fold solution of DFT. 

\medskip

In this paper, we have considered just one example of non-Abelian $T$-duality, 
but it would be interesting to study a lot of examples as a new technique 
to generate GSE solutions. In fact, it is well-known that non-Abelian $T$-duality 
is a systematic method to construct $T$-fold solutions in DFT.

\subsection{YB-deformed $\AdS5\times\rmS^5$ backgrounds}
\label{sec:non-geometry-AdS5xS5}

We show that various YB deformations of the $\AdS5\times\rmS^5$ background are $T$-folds. 
We consider here examples associated with the following five classical $r$-matrices:
\begin{enumerate}
\setlength{\leftskip}{0.7cm} 
\item \quad $r = \frac{1}{2\,\eta}\,\bigl[\eta_1\,(D+M_{+-})\wedge P_+ 
+ \eta_2\,M_{+2}\wedge P_3 \bigr]$\,, 

\item \quad $r = \frac{1}{2}\,P_0\wedge D$\,, 

\item \quad $r = \frac{1}{2}\,\bigl[P_0\wedge D + P^i \wedge (M_{0i}+ M_{1i})\bigr]$\,, 

\item \quad $r = \frac{1}{2\eta}\, P_- \wedge (\eta_1\,D-\eta_2\,M_{+-})$\,, 

\item \quad $r = \frac{1}{2}\, M_{-\mu}\wedge P^\mu$\,. 
\end{enumerate}
The classical $r$-matrices other than the first one are non-unimodular. 
Note here that the $\rmS^5$ part remains undeformed and only the $\AdS5$ part is deformed. 
As shown in App.\,\ref{sec:Penrose-limit}, through the (modified) Penrose limit, 
the second and third examples are reduced to the two examples discussed in the previous subsection. 

\subsubsection{Non-Abelian unimodular $r$-matrix}

Let us consider a non-Abelian unimodular $r$-matrix 
(see $R_5$ in Tab.\,$1$ of \cite{Borsato:2016ose}),
\begin{align}
 r = \frac{1}{2\,\eta}\,\bigl[\eta_1\,(D+M_{+-})\wedge P_+ + \eta_2\,M_{+2}\wedge P_3 \bigr] \,,
\end{align}
where, for simplicity, it is written in terms of the light-cone coordinates,%
\footnote{In the following, our light-cone convention is taken as $\varepsilon_{z+-23r\xi\phi_1\phi_2\phi_3} = + \sqrt{\abs{\CG}}$ rather than \eqref{eq:epsilon}.}
\begin{align}
 x^\pm \equiv \frac{x^0 \pm x^1}{\sqrt{2}} \,. 
\end{align}
The corresponding YB-deformed background is given by 
\begin{align}
 \rmd s^2 &= \frac{\rmd z^2}{z^2} + \frac{z^2\,[(\rmd x^2)^2+(\rmd x^3)^2]}{z^4+(\eta_2\,x^-)^2} 
 - \frac{2\,z^2\,\rmd x^+\,\rmd x^- - 4\,\eta_1^2\,z^{-1}\,x^-\,\rmd z\,\rmd x^-}{z^4- (2\,\eta_1\,x^-)^2}
\nn\\
 &\quad + \frac{2\,\{[x^2 (2\,\eta_1^2+\eta_2^2)-\eta_1\,\eta_2\,x^3]\,z^2\,x^-\,\rmd x^2+\eta_1\,(2\,\eta_1\,x^3-\eta_2\,x^2)\,\rmd x^3\}\,\rmd x^-}{[z^4-(2\,\eta_1\,x^-)^2]\,[z^4+(\eta_2\,x^-)^2]}
\nn\\
 &\quad - \frac{(\eta_1^2+\eta_2^2)\,(z\,x^2)^2 -2\,\eta_1\,\eta_2\,z^2\,x^2\,x^3 + \eta_1^2\,[z^4 + (z\,x^3)^2+ (\eta_2\,x^-)^2]}{[z^4-(2\,\eta_1\,x^-)^2]\,[z^4+(\eta_2\,x^-)^2]}\,(\rmd x^-)^2
 + \rmd s_{\rmS^5}^2\,,
\nn\\
 B_2 &= -\biggl[\frac{\eta_1\,\{x^2\,[z^4+2\,(\eta_2\,x^-)^2]-2\,\eta_1\,\eta_2\,(x^-)^2\,x^3\}\,\rmd x^2+\{\eta_1\,z^4\,x^3-\eta_2\,x^2\,[z^4-2\,(\eta_1\,x^-)^2]\}\,\rmd x^3}{[z^4- (2\,\eta_1\,x^-)^2]\,[z^4+(\eta_2\,x^-)^2]}
\nn\\
 &\qquad +\frac{\eta_1\,(z\,\rmd z - 2\, x^-\,\rmd x^+)}{z^4-(2\,\eta_1\,x^-)^2}\biggr]\wedge \rmd x^-
 +\frac{\eta_2\,x^-\,\rmd x^2 \wedge \rmd x^3}{z^4+(\eta_2\,x^-)^2} \,,
\nn\\
 \Phi &=\frac{1}{2} \ln \biggl[\frac{z^8}{[z^4- (2\,\eta_1\,x^-)^2]\,[z^4+ (\eta_2\,x^-)^2]}\biggr]\,,
\nn\\
 \hat{\sfF}_1 &=\frac{4\,\eta_1\,\eta_2\,x^-\,(2\,x^-\,\rmd z -z\,\rmd x^-)}{z^5} \,,
\nn\\
 \hat{\sfF}_3 &= -B_2\wedge \sfF_1 +\frac{4\,\eta_1}{z^5}\, \bigl(2\,x^-\,\rmd z - z\,\rmd x^-\bigr)\wedge \rmd x^2 \wedge \rmd x^3 
\nn\\
 &\quad +\frac{4}{z^5}\, \rmd z \wedge \rmd x^- \wedge \bigl[\eta_1\,(x^3\,\rmd x^2- x^2\,\rmd x^3) + \eta_2\,(x^-\,\rmd x^+ - x^2\,\rmd x^2) \bigr] \,,
\nn\\
 \hat{\sfF}_5&=4\,\biggl[\frac{z^8}{[z^4-(2\,\eta_1\,x^-)^2]\,[z^4+ (\eta_2\,x^-)^2]}\,\omega_{\AdS5} + \omega_{\rmS^5}\biggr] \,,
\nn\\
 \hat{\sfF}_7&= -B_2\wedge \sfF_5 \,,\qquad 
 \hat{\sfF}_9= -\frac{1}{2}\,B_2\wedge \sfF_7 \,. 
\end{align}
In terms of the dual fields, we obtain the following expression:
\begin{align}
\begin{split}
 \rmd s_{\text{dual}}^2 &= \frac{\rmd z^2 -2\,\rmd x^+\,\rmd x^-+ (\rmd x^2)^2
 +(\rmd x^3)^2}{z^2} + \rmd s_{\rmS^5}^2\,, \qquad 
 \tilde{\phi}=0\,, \\
 \beta &=\eta_1\,\bigl(z\,\partial_z+2\,x^-\,\partial_-+x^2\,\partial_2+x^3\,\partial_3\bigr)
 \wedge\partial_+ + \eta_2\,\bigl(x^2\,\partial_+ + x^-\,\partial_2\bigr)\wedge \partial_3 \,.
\end{split}
\end{align}
It is straightforward to check that the R-R field strengths are given by 
\begin{align}
 \hat{\sfF} = \Exp{-B_2\wedge}\sfF\,,\qquad \sfF \equiv \Exp{-\beta\vee}\check{\sfF}\,,\qquad 
 \check{\sfF} = 4\,\bigl(\omega_{\AdS5} + \omega_{\rmS^5}\bigr)\,.
\end{align}
Namely, as advocated in Sec.\,\ref{sec:YB-beta-deform}, 
the $\beta$-untwisted R-R fields $\check{\sfF}$ are invariant 
under the YB deformation. 

\medskip

This background has the following components of $Q$-flux:
\begin{align}
 Q_z{}^{z+} = \eta_1\,,\quad 
 Q_-{}^{-+} = 2\,\eta_1\,,\quad 
 Q_2{}^{2+} = \eta_1\,,\quad 
 Q_3{}^{3+} = \eta_1\,,\quad 
 Q_2{}^{+3} = \eta_2\,,\quad 
 Q_-{}^{23} = \eta_2\,. 
\end{align}
Accordingly, for example, when the $x^3$ direction is compactified with a period 
$x^3\sim x^3+\eta_1^{-1}$\,, this background becomes a $T$-fold with the monodromy,
\begin{align}
 \cH_{MN}(x^3 +\eta_1^{-1}) = \bigl[\Omega^\rmT\cH(x)\,\Omega\bigr]_{MN} \,,\qquad 
 \Omega^M{}_N\equiv \begin{pmatrix} \delta^m_n & 2\,\delta_3^{[m}\,\delta_+^{n]} \\ 0 & \delta_m^n \end{pmatrix} \in \OO(10,10;\mathbb{Z})\,. 
\end{align}
The R-R fields $\sfF$ are also twisted by the same monodromy,
\begin{align}
 \sfF(x^3+\eta_1^{-1}) = \Exp{- \omega\vee}\sfF(x^3)\,,\qquad 
 \omega^{mn} = 2\,\delta_3^{[m} \,\delta_+^{n]} \,. 
\end{align}
Note that the R-R potentials are twisted by the same monodromy as well, 
though their explicit forms are not written down here. 

\subsubsection{$r=\frac{1}{2}\,P_0\wedge D$}
\label{sec:AdS-P0-D}

Let us next consider a classical $r$-matrix\cite{vanTongeren:2015uha,Orlando:2016qqu},
\begin{align}
 r=\frac{1}{2}\,P_0\wedge D\,.
\end{align}
Because $[P_0, D] \neq 0$\,, this classical $r$-matrix does not satisfy the unimodularity condition. 
By introducing the polar coordinates,
\begin{align}
 x^1 =\rho\sin\theta \cos\phi\,, \qquad x^2 =\rho\sin\theta\sin\phi\,, 
 \qquad x^3 =\rho\cos\theta\,,
\end{align}
the deformed background can be rewritten as \cite{Orlando:2016qqu}\footnote{Only the metric 
and NS-NS two-form were computed in \cite{vanTongeren:2015uha}.}
\begin{align}
\label{eq:AdS-P0-D}
\begin{split}
 \rmd s^2 &= \frac{z^2\,\bigl[\rmd z^2 -(\rmd x^0)^2+\rmd\rho^2\bigr]-\eta^2\,(\rmd \rho -\rho\,z^{-1}\,\rmd z)^2}{z^4-\eta^2\,(z^2+\rho^2)}
    +\frac{\rho^2\,(\rmd \theta^2+\sin^2\theta\,\rmd \phi^2)}{z^2}
    +\rmd s_{\rmS^5}^2 \,, 
\\
    B_2 &= -\eta\,\frac{\rmd x^0 \wedge (z\,\rmd z + \rho\,\rmd \rho)}{z^4-\eta^2\,(z^2+\rho^2)}\,, \qquad 
 \Phi = \frac{1}{2}\ln \biggl[\frac{z^4}{z^4-\eta^2\,(z^2+\rho^2)}\biggr]\,,\qquad 
 I = -\eta\,\partial_0 \,,
\\
 \hat{\sfF}_1 &=0\,,\quad 
 \hat{\sfF}_3 = \frac{4\,\eta\,\rho^2\sin\theta}{z^5}\, (z\,\rmd \rho - \rho \, \rmd z)\wedge \rmd \theta\wedge \rmd \phi \,,
\\
 \hat{\sfF}_5 &= 4\,\biggl[\frac{z^4}{z^4-\eta^2\, (z^2+\rho^2)}\,\omega_{\AdS5} + \omega_{\rmS^5}\biggr]\,,
\\
 \hat{\sfF}_7&=\frac{4\,\eta\,\rmd x^0 \wedge (z\,\rmd z+\rho\,\rmd \rho)}{z^4-\eta^2\, (z^2+\rho^2)}\wedge \omega_{\rmS^5}\,,\quad \hat{\sfF}_9=0\,. 
\end{split}
\end{align}
This background is not a solution of the usual type IIB supergravity, 
but that of GSE \cite{Arutyunov:2015mqj}. 
By setting $\eta=0$, this background reduces to the original $\AdS5\times \rmS^5$. 

\medskip

In the dual parameterization, the dual metric, the $\beta$ field and the dual dilaton 
are given by 
\begin{align}
\begin{split}
 \rmd s_{\text{dual}}^2 &= \frac{\rmd z^2 -(\rmd x^0)^2+(\rmd x^1)^2+(\rmd x^2)^2+(\rmd x^3)^2}{z^2} +\rmd s_{\rmS^5}^2 \,,\qquad \tilde{\phi}=0\,,
\\
 \beta &= \eta\,\hat{P}_0\wedge \hat{D}=\eta\,\partial_0\wedge (z\,\partial_z + x^1\,\partial_1 + x^2\,\partial_2 + x^3\,\partial_3) 
 = \eta\,\partial_0\wedge (z\,\partial_z + \rho\,\partial_\rho)\,. 
\end{split}
\end{align}
The Killing vector $I^m$ satisfies the divergence formula,
\begin{align}
 I^0 = -\eta = \tilde{D}_{m}\beta^{0m} \,. 
\end{align}
The $Q$-flux has the following non-vanishing components:
\begin{align}
 Q_z{}^{0z} = Q_1{}^{01} = Q_2{}^{02} = Q_3{}^{03} = \eta\,. 
\end{align}
Thus, when at least one of the $(x^1,x^2,x^3)$ directions is compactified, 
the background can be interpreted as a $T$-fold. 
For example, when the $x^1$ direction is compactified, the monodromy is given by 
\begin{align}
 \cH_{MN}(x^1+\eta^{-1}) = \bigl[\Omega^\rmT\cH(x^1)\,\Omega\bigr]_{MN} \,,\qquad 
 \Omega^M{}_N \equiv \begin{pmatrix} \delta^m_n & 2\,\delta_0^{[m} \,\delta_1^{n]} \\ 0 & \delta_m^n \end{pmatrix}\,. 
\label{eq:monodromy-AdS-P0-D}
\end{align}

\medskip

From \eqref{eq:AdS-P0-D}, the R-R potentials can be found as follows:
\begin{align}
\begin{split}
 \hat{\sfC}_0&=0\,,\quad 
 \hat{\sfC}_2= \frac{\eta\,\rho^3\sin\theta}{z^4}\,\rmd \theta\wedge \rmd \phi\,, 
\\
 \hat{\sfC}_4&= \frac{\rho^2\sin\theta}{z^4}\,\rmd x^0\wedge \rmd \rho\wedge \rmd \theta \wedge \rmd\phi + \omega_4 -B_2 \wedge \hat{\sfC}_2 \,,
\\
 \hat{\sfC}_6&= -B_2\wedge \omega_4 \,, \quad
 \hat{\sfC}_8 = 0 \,. 
\end{split}
\end{align}
Providing the $B$-twist, we obtain
\begin{align}
\begin{split}
 \sfF_1&=0\,,\quad 
 \sfF_3 = \frac{4\,\eta \,\rho^2 \sin \theta}{z^5}\, (\rho\,\rmd z- z\,\rmd \rho) \wedge \rmd \theta \wedge \rmd \phi\,,
\\
 \sfF_5&=4\,\bigl(\omega_{\AdS5}+\omega_{\rmS^5} \bigr)\,,\quad \sfF_7 =0\,,\quad \sfF_9 =0\,,
\\
 \sfA_0&=0\,,\quad \sfA_2 = \frac{\eta\,\rho^3 \sin \theta}{z^4}\,\rmd \theta \wedge\rmd \phi\,,
\\
 \sfA_4&= \frac{\rho^2 \sin\theta}{z^4}\,\rmd x^0 \wedge \rmd \rho \wedge \rmd \theta \wedge \rmd \phi+\omega_4 \,,\quad \sfA_6=0\,,\quad \sfA_8=0\,. 
\end{split}
\end{align}
We can further compute the $\beta$-untwisted fields,
\begin{align}
\begin{split}
 \check{\sfF}_1 &=0\,,\quad 
 \check{\sfF}_3 = 0\,, \quad
 \check{\sfF}_5 =4\,\bigl(\omega_{\AdS5}+\omega_{\rmS^5}\bigr)\,,\quad \check{\sfF}_7 =0\,,\quad \check{\sfF}_9 =0\,,
\\
 \check{\sfC}_0 &=0\,,\quad \check{\sfC}_2 = 0\,, \quad
 \check{\sfC}_4 = \frac{\rho^2 \sin\theta}{z^4}\,\rmd x^0 \wedge \rmd \rho \wedge \rmd \theta \wedge \rmd \phi +\omega_4 \,,\quad \check{\sfC}_6=0\,,\quad \check{\sfC}_8=0\,.
\end{split}
\end{align}
As expected, the $\beta$-untwisted R-R fields are precisely the R-R fields in the undeformed background, and they are single-valued. 
In terms of the twisted R-R fields, $(\sfF,\,\sfA)$, the R-R fields have the same monodromy as \eqref{eq:monodromy-AdS-P0-D},
\begin{align}
 \sfA(x^1+\eta^{-1})= \Exp{- \omega \vee} \sfA(x^1)\,,\quad \sfF(x^1+\eta^{-1}) = \Exp{- \omega \vee}\sfF(x^1)\,,\quad 
 \omega^{mn} = 2\,\delta_0^{[m} \,\delta_1^{n]} \,. 
\end{align}

\subsubsection{A scaling limit of the Drinfeld--Jimbo $r$-matrix}
\label{sec:AdS-Drinfeld--Jimbo}

Let us consider a classical $r$-matrix \cite{Hoare:2016hwh,Orlando:2016qqu},
\begin{align}
 r = \frac{1}{2}\,\bigl[P_0\wedge D + P^i \wedge (M_{0i}+ M_{1i})\bigr] \,,
\end{align}
which can be obtained as a scaling limit of the classical $r$-matrix of 
Drinfeld-Jimbo type \cite{Drinfeld:1985rx,Jimbo:1985zk}. 
By using the polar coordinates $(\rho,\theta)$,
\begin{align}
 (\rmd x^2)^2 + (\rmd x^3)^2 = \rmd \rho^2+\rho^2\,\rmd \theta^2 \,,
\end{align}
the YB-deformed background, which satisfies GSE, is given by \cite{Hoare:2016hwh,Orlando:2016qqu}
\begin{align}
 \rmd s^2 &= \frac{\rmd z^2-(\rmd x^0)^2}{z^2- \eta^2}
    +\frac{z^2\bigl[(\rmd x^1)^2+\rmd \rho^2\bigr]}{z^4+ \eta^2\,\rho^2}
    +\frac{\rho^2\,\rmd \theta^2}{z^2}+ \rmd s^2_{\rmS^5}\,, 
\nn\\
 B_2 &= \eta\,\biggl[\frac{\rmd z \wedge \rmd x^0}{z\,(z^2-\eta^2)} - \frac{\rho\,\rmd x^1 \wedge \rmd \rho}{z^4+ \eta^2\,\rho^2}\biggr]\,, 
\nn\\
 \Phi &= \frac{1}{2}\ln\biggl[\frac{z^6}{(z^2- \eta^2)(z^4+ \eta^2\,\rho^2)}\biggr]\,,\qquad 
 I = -\eta\,(4\,\partial_0 + 2\,\partial_1) \,, 
\nn\\
 \hat{\sfF}_1&= -\frac{4\,\eta^2\,\rho^2}{z^4}\, \rmd\theta\,, \quad 
 \hat{\sfF}_3 = 4\,\eta\,\rho\,\biggl(\frac{\rho\,\rmd z \wedge \rmd x^0}{z\,(z^4-\eta^2\,z^2)} + \frac{\rmd x^1 \wedge \rmd \rho}{z^4+ \eta^2\,\rho^2}\biggr)\wedge \rmd \theta \,,
\nn\\
 \hat{\sfF}_5&= 4\,\biggl[\frac{z^6}{(z^2- \eta^2) (z^4 + \eta^2\, \rho^2)}\,\omega_{\AdS5} +\omega_{\rmS^5}\biggr]\,,
\nn\\
 \hat{\sfF}_7&=4\,\eta\,\biggl(-\frac{\rmd z \wedge \rmd x^0}{z\,(z^2- \eta^2)}+\frac{\rho\,\rmd x^1 \wedge \rmd \rho}{z^4+ \eta^2\,\rho^2}\biggr) \wedge \omega_{\rmS^5} \,,
\nn\\
 \hat{\sfF}_9&=-\frac{4\,\eta^2\,\rho}{z\,(z^2- \eta^2) (z^4+ \eta^2\,\rho^2)}\,\rmd z \wedge \rmd x^0 \wedge \rmd x^1\wedge \rmd \rho \wedge\omega_{\rmS^5}\,. 
\label{eq:HvT}
\end{align}
The R-R potentials can be found as follows:
\begin{align}
\begin{split}
 \hat{\sfC}_0&= 0\,, \quad
 \hat{\sfC}_2 = -\frac{\eta\,\rho^2}{z^4}\,\rmd x^0\wedge \rmd \theta\,,
\quad
 \hat{\sfC}_4 = \frac{\rho}{z^4+ \eta^2\,\rho^2}\, \rmd x^0\wedge \rmd x^1\wedge \rmd\rho \wedge \rmd\theta + \omega_4 \,,
\\
 \hat{\sfC}_6&= -B_2\wedge \omega_4 \,, \quad
 \hat{\sfC}_8 = \frac{\eta^2\,\rho}{z\,(z^2-\eta^2)(z^4+\eta^2\,\rho^2)}\,\rmd x^0\wedge \rmd x^1\wedge \rmd\rho\wedge \rmd z\wedge \omega_4 \,.
\end{split}
\end{align}

\medskip

Then the corresponding dual fields in the NS-NS sector are given by 
\begin{align}
\begin{split}
 \rmd s_{\text{dual}}^2 &= \frac{\rmd{z}^2 - (\rmd x^0)^2+(\rmd x^1)^2+\rmd \rho^2+\rho^2\,\rmd \theta^2}{z^2} + \rmd s^2_{\rmS^5}\,,\qquad \tilde{\phi}=0\,, 
\\
 \beta &= \eta\,\bigl[\hat{P}_0\wedge \hat{D}+\hat{P}^i\wedge (M_{0i}+ M_{1i})\bigr]
 = \eta\,(z\,\partial_0\wedge \partial_z - x^2\,\partial_1\wedge\partial_2 - x^3\,\partial_1\wedge\partial_3)
\nn\\
 &=\eta\,(z\,\partial_0\wedge \partial_z -\rho\,\partial_1\wedge\partial_\rho) \,, 
\end{split}
\end{align}
and the Killing vector $I^m$ again satisfies the divergence formula,
\begin{align}
 I^0 = -4\,\eta = \tilde{D}_m\beta^{0m} \,,\qquad I^1 = -2\,\eta = \tilde{D}_m\beta^{1m} \,.
\end{align}

Providing the $B$-twist to the R-R field strengths, we obtain
\begin{align}
\begin{split}
 \sfF_1&=-\frac{4\,\eta^2 \rho^2}{z^4}\,\rmd \theta \,,\quad 
 \sfF_3 = \frac{4\,\eta\,\rho}{z^5}\, \bigl(\rho\,\rmd z\wedge \rmd x^0 + z\,\rmd x^1\wedge \rmd\rho\bigr) \wedge \rmd \theta \,,
\\
 \sfF_5&=4\,\bigl(\omega_{\AdS5}+\omega_{\rmS^5} \bigr)\,,\quad \sfF_7 =0\,,\quad \sfF_9 =0\,,
\\
 \sfA_0 &=0\,,\quad 
 \sfA_2 = -\frac{\eta\,\rho^2}{z^4}\, \rmd x^0\wedge \rmd \theta \,,
\\
 \sfA_4 &= \frac{\rho}{z^4} \, \rmd x^0 \wedge \rmd x^1 \wedge \rmd \rho \wedge \rmd \theta +\omega_4 \,, \quad
 \sfA_6 =0\,,\quad 
 \sfA_8 =0\,. 
\end{split}
\end{align}
Furthermore, the $\beta$-untwist leads to the following expressions: 
\begin{align}
\begin{split}
 \check{\sfF}_1 &=0\,,\quad 
 \check{\sfF}_3 = 0\,, \quad
 \check{\sfF}_5 =4\,\bigl(\omega_{\AdS5}+\omega_{\rmS^5}\bigr)\,,\quad \check{\sfF}_7 =0\,,\quad \check{\sfF}_9 =0\,,
\\
 \check{\sfC}_0 &=0\,,\quad \check{\sfC}_2 = 0\,, \quad
 \check{\sfC}_4 = \frac{\rho}{z^4} \, \rmd x^0 \wedge \rmd x^1 \wedge \rmd \rho \wedge \rmd \theta +\omega_4 \,,\quad \check{\sfC}_6=0\,,\quad \check{\sfC}_8=0\,. 
\end{split}
\end{align}
These are the same as the undeformed R-R potentials. 

\medskip

Then the non-zero component of $Q$-flux are given by 
\begin{align}
 Q_z{}^{0z} = \eta\,,\qquad Q_2{}^{12} = -\eta\,,\qquad Q_3{}^{13} = -\eta\,. 
\end{align}
When the $x^2$-direction is compactified as $x^2 \sim x^2 +\eta^{-1}$, 
this background becomes a $T$-fold with the monodromy,
\begin{align}
\begin{split}
 \cH_{MN}(x^2+\eta^{-1}) &= \bigl[\Omega^\rmT\cH(x^2)\,\Omega\bigr]_{MN} \,,\qquad 
 \Omega^M{}_N \equiv \begin{pmatrix} \delta^m_n & -2\,\delta_1^{[m} \,\delta_2^{n]} \\ 0 & \delta_m^n \end{pmatrix}\,,
\\
 \sfF(x^2+\eta^{-1}) &= \Exp{- \omega \vee}\sfF(x^2)\,,\qquad 
 \omega^{mn} = -2\,\delta_1^{[m} \,\delta_2^{n]} \,. 
\end{split}
\end{align}

\subsubsection{$r=\frac{1}{2\,\eta}\,P_-\wedge (\eta_1\,D-\eta_2\,M_{+-})$}

Let us consider a non-unimodular $r$-matrix\footnote{This $r$-matrix includes the known examples studied in Sec.\,4.3 ($\eta_1=-\eta_2=-\eta$) 
and 4.4 ($\eta_1=-\eta$, $\eta_2=0$) of \cite{Orlando:2016qqu} as special cases. },
\begin{align}
 r=\frac{1}{2\,\eta}\,P_-\wedge (\eta_1\,D-\eta_2\,M_{+-})\,. 
\end{align}
Here we have introduced the light-cone coordinates and polar coordinates as
\begin{align}
 x^\pm \equiv \frac{x^0 \pm x^1}{\sqrt{2}} \,, \qquad 
 (\rmd x^2)^2 + (\rmd x^3)^2 = \rmd \rho^2+\rho^2\,\rmd \theta^2 \,. 
\end{align}

\medskip 

The YB-deformed background is given by
\begin{align}
 \rmd s^2&= \frac{\rmd z^2 + \rmd \rho^2 + \rho^2\,\rmd \theta^2}{z^2}
 - \frac{2\, z^2\,\rmd x^+\,\rmd x^-}{z^4 - (\eta_1 + \eta_2)^2\,(x^+)^2} 
\nn\\
 &\quad + \eta_1\,\rmd x^+\,\frac{2\,x^+\,(\eta_1 + \eta_2)\,(z\,\rmd z + \rho\,\rmd \rho) - \eta_1\,(z^2 + \rho^2)\,\rmd x^+}{z^2\,[z^4 - (\eta_1 + \eta_2)^2\,(x^+)^2]} + \rmd s^2_{\rmS^5}\,,
\nn\\
 B_2 &= \eta_1\,\frac{x^+\,\rmd x^+\wedge \rmd x^- + z\,\rmd z\wedge \rmd x^+ - \rho\,\rmd x^+\wedge \rmd \rho}{z^4 - (\eta_1 + \eta_2)^2\,(x^+)^2}
     + \eta_2\,\frac{x^+\,\rmd x^+\wedge \rmd x^-}{z^4 - (\eta_1 + \eta_2)^2\,(x^+)^2}\,,
\nn\\
 \Phi &= \frac{1}{2} \ln\biggl[\frac{z^4}{z^4 - (\eta_1 + \eta_2)^2\, (x^+)^2}\biggr]\,,\qquad 
 I = -(\eta_1 - \eta_2)\,\partial_-\,,
\nn\\
 \hat{\sfF}_1&= 0\,,\quad 
 \hat{\sfF}_3 = -\frac{4\,\rho\,\bigl[\eta_1\,(\rho\, \rmd z\wedge \rmd x^+ + z\,\rmd x^+\wedge \rmd \rho - x^+\,\rmd z\wedge \rmd \rho)
                                    - \eta_2\,x^+\,\rmd z\wedge \rmd \rho\bigr]\wedge \rmd \theta}{z^5}\,,
\nn\\
 \hat{\sfF}_5&= 4 \biggl[\frac{z^4}{z^4 - (\eta_1 + \eta_2)^2\,(x^+)^2}\,\omega_{\AdS5} + \omega_{\rmS^5}\biggr] \,,
\nn\\
 \hat{\sfF}_7&= -\frac{4\,\bigl[\eta_1\,(x^+\,\rmd x^+\wedge \rmd x^- + z\,\rmd z\wedge \rmd x^+ - \rho\,\rmd x^+\wedge \rmd \rho) + \eta_2\,x^+\,\rmd x^+ \wedge \rmd x^-\bigr]\wedge \omega_{\rmS^5}}{z^4 - (\eta_1+\eta_2)^2\,(x^+)^2} \,,
\nn\\
 \hat{\sfF}_9&= 0\,. 
\end{align}
The R-R potentials are also given by
\begin{align}
\begin{split}
 \hat{\sfC}_0&= 0\,,\qquad
 \hat{\sfC}_2 = \frac{\rho\,[\eta_1\,\rho\,\rmd x^+ - (\eta_1 + \eta_2)\,x^+\,\rmd \rho]\wedge \rmd \theta}{z^4}\,,
\\
 \hat{\sfC}_4&= \frac{\rho\,\rmd x^+\wedge [z^3\,\rmd x^- - \eta_1\,(\eta_1 + \eta_2)\,x^+\,\rmd z]\wedge \rmd\rho \wedge \rmd\theta}{z^3\,[z^4 - (\eta_1 + \eta_2)^2\,(x^+)^2]} + \omega_4\,,
\\
 \hat{\sfC}_6&= -B_2\wedge \omega_4\,,\qquad 
 \hat{\sfC}_8 = 0\,.
\end{split}
\end{align}

\medskip

The dual fields are given by
\begin{align}
\begin{split}
 \rmd s_{\text{dual}}^2&= \frac{\rmd z^2 -2\,\rmd x^+\,\rmd x^- + \rmd \rho^2 + \rho^2\,\rmd \theta^2}{z^2} + \rmd s^2_{\rmS^5} \,,\qquad 
 \tilde{\phi}=0 \,,
\\
 \beta &= \hat{P}_-\wedge (\eta_1\,\hat{D}+\eta_2\,\hat{M}_{+-}) = \eta_1 \, \partial_- \wedge (z\,\partial_z + x^+\,\partial_+ + \rho \,\partial_\rho)
         +\eta_2 \,x^+\, \partial_- \wedge \partial_+ 
\\
 &= \eta_1 \, \partial_- \wedge (z\,\partial_z + x^+\,\partial_+ + x^2\,\partial_2 + x^3\,\partial_3)
         +\eta_2 \,x^+\, \partial_- \wedge \partial_+ \,,
\end{split}
\end{align}
and the $Q$-flux has the following non-vanishing components:
\begin{align}
 Q_z{}^{-z} = Q_+{}^{-+} = Q_2{}^{-2} = Q_3{}^{-3} = \eta_1\,,\qquad 
 Q_+{}^{-+} = \eta_2 \,. 
\end{align}
In a similar manner as the previous examples, by compactifying one of the $x^1$, $x^2$, and $x^3$ directions with a certain period, this background can also be regarded as a $T$-fold. 
For example, if we make the identification, $x^3\sim x^3 + \eta_1^{-1}$, 
the associated monodromy becomes
\begin{align}
\begin{split}
 \cH_{MN}(x^3+\eta_1^{-1}) &= \bigl[\Omega^\rmT\cH(x^3)\,\Omega\bigr]_{MN} \,,\qquad 
 \Omega^M{}_N \equiv \begin{pmatrix} \delta^m_n & 2\,\delta_-^{[m} \,\delta_3^{n]} \\ 0 & \delta_m^n \end{pmatrix}\,,
\\
 \sfF(x^3+\eta_1^{-1}) &= \Exp{- \omega \vee}\sfF(x^3)\,,\qquad 
 \omega^{mn} = 2\,\delta_-^{[m} \,\delta_3^{n]} \,. 
\end{split}
\end{align}

\paragraph{A solution of Generalized Type IIA Supergravity Equations:}

In the background \eqref{eq:HvT}, by performing a $T$-duality along the $x^1$-direction (see \cite{Sakamoto:2017wor} for the duality transformation rule), 
we obtain the following solution of the generalized type IIA equations of motion:
\begin{align}
 \rmd s^2 &= \frac{\rmd z^2 - (\rmd x^0)^2}{z^2 - \eta^2} + z^2\,(\rmd x^1)^2 
 + \frac{(\rmd \rho + \eta\, \rho\,\rmd x^1)^2 + \rho^2\,\rmd \theta^2}{z^2} + \rmd s_{\rmS^5} \,,
\nn\\
 B_2 &= \frac{\eta\, \rmd z \wedge \rmd x^0}{z\,(z^2 - \eta^2)}\,,\qquad 
 \Phi = -2\,\eta\,x^1 - \frac{1}{2} \ln\Bigl(\frac{z^2 - \eta^2}{z^4}\Bigr)\,,\qquad I=-4\,\eta\,\partial_0\,,
\nn\\
 \hat{\sfF}_2&= \frac{4\,\eta\Exp{2\,\eta\,x^1}\rho\,(\rmd\rho + \eta\,\rho\,\rmd x^1)\wedge \rmd \theta}{z^4} \,,
\nn\\
 \hat{\sfF}_4&= - \frac{4\Exp{2\,\eta\,x^1}\rho\,\rmd z \wedge \rmd x^0 \wedge (\rmd \rho+\eta\,\rho\,\rmd x^1)\wedge \rmd \theta}{z^3\,(z^2- \eta^2)} \,,
\nn\\
 \hat{\sfF}_6&=-4 \Exp{2\,\eta\,x^1} \rmd x^1 \wedge \omega_{\rmS^5}\,,\qquad
 \hat{\sfF}_8 = \frac{4\,\eta\Exp{2\,\eta\,x^1}\rmd z \wedge \rmd x^0 \wedge \rmd x^1 \wedge \omega_{\rmS^5}}{z\,(z^2 - \eta^2)} \,. 
\end{align}
Here the R-R potentials are given by 
\begin{align}
\begin{split}
 \hat{\sfC}_1 &= 0 \,, \qquad
 \hat{\sfC}_3 = \Exp{2\,\eta\,x^1} \frac{\rho\,\rmd x^0 \wedge (\rmd \rho + \eta\,\rho\, \rmd x^1)\wedge \rmd \theta}{z^4} \,,
\\
 \hat{\sfC}_5 &= \Exp{2\,\eta\,x^1} \rmd x^1 \wedge \omega_4 \,, \qquad
 \hat{\sfC}_7 = -\Exp{2\,\eta\,x^1} \frac{\eta\,\rmd z\wedge \rmd x^0 \wedge \rmd x^1 \wedge \omega_4}{z\,(z^2- \eta^2)} \,.
\end{split}
\end{align}
This background cannot be regarded as a $T$-fold, 
but it is the first example of the solution for the generalized type IIA supergravity equations.

\subsubsection{$r = \frac{1}{2}\, M_{-\mu}\wedge P^\mu$}

The final example is associated with the $r$-matrix\cite{Orlando:2016qqu}
\begin{align}
 r = \frac{1}{2}\, M_{-\mu}\wedge P^\mu\,.
\end{align}
This $r$-matrix is called the light-cone $\kappa$-Poincar\'e.
Again, by introducing the coordinates,
\begin{align}
 x^\pm \equiv \frac{x^0 \pm x^1}{\sqrt{2}} \,, \qquad 
 (\rmd x^2)^2 + (\rmd x^3)^2 = \rmd \rho^2+\rho^2\,\rmd \theta^2 \,,
\end{align}
the YB-deformed background is given by (see Sec.\,4.5 of \cite{Orlando:2016qqu})
\begin{align}
 \rmd s^2&=\frac{\frac{z^4}{z^4 - (\eta\,x^+)^2}\,(\rmd z^2 -2\,\rmd x^+\,\rmd x^-) - \eta^2 \,\frac{(x^+\,\rmd z)^2 + (\rho \,\rmd x^+)^2 -2\,x^+ \rho\,\rmd x^+\,\rmd \rho}{z^4 - (\eta\,x^+)^2} + \rmd \rho^2 + \rho^2\,\rmd \theta^2}{z^2} 
 + \rmd s_{\rmS^5}^2 \,,
\nn\\
 B_2 &= \frac{\eta\,\rmd x^+\wedge (x^+\,\rmd x^- -\rho\,\rmd\rho)}{z^4-(\eta \,x^+)^2}\,,\qquad 
 \Phi =\frac{1}{2} \ln\biggl[\frac{z^4}{z^4-(\eta\,x^+)^2}\biggr] \,,\qquad 
 I^- = 3\,\eta\,,
\nn\\
 \hat{\sfF}_1&=0 \,,\quad
 \hat{\sfF}_3= -\frac{4\,\eta\,\rho}{z^5}\, \rmd z \wedge \bigl(\rho\,\rmd x^+ - x^+\,\rmd \rho\bigr)\wedge \rmd\theta \,,\quad
 \hat{\sfF}_5= 4\,\biggl[\frac{z^4}{z^4-(\eta\,x^+)^2}\,\omega_{\AdS5}+\omega_{\rmS^5} \biggr] \,,
\nn\\
 \hat{\sfF}_7&= - \frac{4\,\eta}{z^4-(\eta\,x^+)^2}\,\rmd x^+ \wedge \bigl(x^+\,\rmd x^- - \rho\,\rmd \rho\bigr)\wedge \omega_{\rmS^5} \,,\qquad
 \hat{\sfF}_9= 0\,. 
\end{align}
The R-R potentials can be found as follows:
\begin{align}
\begin{split}
 \hat{\sfC}_0&=0\,,\quad 
 \hat{\sfC}_2= \frac{\eta\,\rho}{z^4}\,\bigl(\rho\,\rmd x^+-x^+\,\rmd \rho\bigr)\wedge\rmd \theta \,,
\\
 \hat{\sfC}_4& =\frac{\rho}{z^4-(\eta\,x^+)^2}\, \rmd x^+\wedge \rmd x^-\wedge \rmd\rho \wedge \rmd\theta + \omega_4\,, \quad 
 \hat{\sfC}_6 = -B_2\wedge \omega_4 \,, \quad 
 \hat{\sfC}_8=0 \,.
\end{split}
\end{align}

\medskip

The corresponding dual fields are given by
\begin{align}
\begin{split}
 \rmd s_{\text{dual}}^2
 &= \frac{\rmd z^2 -2\rmd x^+\,\rmd x^- + \rmd \rho^2 + \rho^2\,\rmd \theta^2}{z^2} + \rmd s^2_{\rmS^5} \,, \qquad \tilde{\phi}=0 \,,
\\
 \beta &= \eta\, \hat{M}_{-\mu}\wedge \hat{P}^\mu = \eta\,\partial_-\wedge (x^+\,\partial_+ + \rho\,\partial_\rho) 
 = \eta\,\partial_-\wedge (x^+\,\partial_+ + x^2\,\partial_2 + x^3\,\partial_3) \,,
\end{split}
\end{align}
and it is easy to check that the divergence formula is satisfied: 
\begin{align}
 I^- = 3\,\eta = \tilde{D}_m \beta^{-m} \,.
\end{align}

\medskip 

We can calculate other types of the R-R field fields as
\begin{align}
\begin{split}
 \sfF_1&=0\,,\quad 
 \sfF_3 =-\frac{4\,\eta\,\rho}{z^5}\, \rmd z \wedge (\rho\,\rmd x^+ -x^+\,\rmd \rho)\wedge \rmd \theta \,,
\\
 \sfF_5&=4\,\bigl(\omega_{\AdS5}+\omega_{\rmS^5} \bigr)\,,\quad \sfF_7 =0\,,\quad \sfF_9 =0\,,
\\
 \sfA_0 &=0\,,\quad 
 \sfA_2 = \frac{\eta\,\rho}{z^4}\,\bigl(\rho\,\rmd x^+ -x^+\,\rmd \rho\bigr)\wedge \rmd \theta \,,
\\
 \sfA_4 &= \frac{\rho}{z^4} \, \rmd x^+ \wedge \rmd x^- \wedge \rmd \rho \wedge \rmd \theta +\omega_4 \,, \quad
 \sfA_6 =0\,,\quad 
 \sfA_8 =0\,,
\end{split}
\end{align}
and
\begin{align}
\begin{split}
 \check{\sfF}_1 &=0\,,\quad 
 \check{\sfF}_3 = 0\,, \quad
 \check{\sfF}_5 =4\,\bigl(\omega_{\AdS5}+\omega_{\rmS^5}\bigr)\,,\quad \check{\sfF}_7 =0\,,\quad \check{\sfF}_9 =0\,,
\\
 \check{\sfC}_0 &=0\,,\quad \check{\sfC}_2 = 0\,, \quad
 \check{\sfC}_4 =\frac{\rho}{z^4} \, \rmd x^+ \wedge \rmd x^-\wedge \rmd \rho \wedge \rmd \theta +\omega_4 \,,\quad \check{\sfC}_6=0\,,\quad \check{\sfC}_8=0\,,
\end{split}
\end{align}
and the $\beta$-twisted fields are again invariant under the YB deformation. 

\medskip 

The non-geometric $Q$-flux has the non-vanishing components,
\begin{align}
 Q_+{}^{-+}=Q_2{}^{-2}=Q_3{}^{-3}= \eta \,,
\end{align}
and again by compactifying one of the $x^1$, $x^2$, and $x^3$ directions, 
this background becomes a $T$-fold. 
Namely, if we compactify the $x^3$-direction as, $x^3\sim x^3 + \eta^{-1}$, 
the associated monodromy becomes
\begin{align}
\begin{split}
 \cH_{MN}(x^3+\eta^{-1}) &= \bigl[\Omega^\rmT\cH(x^3)\,\Omega\bigr]_{MN} \,,\qquad 
 \Omega^M{}_N \equiv \begin{pmatrix} \delta^m_n & 2\,\delta_-^{[m} \,\delta_3^{n]} \\ 0 & \delta_m^n \end{pmatrix}\,,
\\
 \sfF(x^3+\eta^{-1}) &= \Exp{- \omega \vee}\sfF(x^3)\,,\qquad 
 \omega^{mn} = 2\,\delta_-^{[m} \,\delta_3^{n]} \,. 
\end{split}
\end{align}

\section{Conclusion and discussion}
\label{sec:discussion}

In this paper, we have first reviewed the notion of $T$-folds by showing two examples: (1) a toy model which shows how to obtain a $T$-fold background upon a chain of dualizations of a geometric torus and (2) the (co-dimension 1) exotic $5_2^2$-brane background.
These $T$-folds require the full set of $T$-duality transformations as transition functions to be globally well-defined. 

\medskip

Then, we have elucidated that the simple formula \eqref{YB-formula} proposed in 
\cite{Sakamoto:2017cpu} and the divergence formula \eqref{eq:div-formula} reproduce 
various YB-deformed backgrounds. This means that the YB deformation with 
a classical $r$-matrix $r=\frac{1}{2}\,r^{ij}\,T_i\wedge T_j$ satisfying the (homogeneous) CYBE, 
is equivalent to the $\beta$-deformation with the deformation parameter 
\begin{align}
 r^{mn} = 2\,\eta\,r^{ij}\, \hat{T}^{m}_i\,\hat{T}_j^{n}\,.
\end{align}
We also considered a known background obtained by the non-Abelian $T$-duality 
and showed that the extra vector $I$ determined by the divergence formula \eqref{eq:div-formula} 
makes the background a solution of GSE. 

\medskip

We have then computed monodromy matrices for various YB-deformed backgrounds 
and a non-Abelian $T$-dual background. 
In order to clarify the general pattern, let us consider a YB deformation associated 
with a classical $r$-matrix,
\begin{align}
 r = \frac{1}{2}\,\bigl[a^{\mu\nu\rho}\,M_{\mu\nu}\wedge P_\rho + b^\mu\, D\wedge P_\mu\bigr] 
\qquad (a^{\mu\nu\rho}=a^{[\mu\nu]\rho},\,b^\mu : \text{ constant})\,,
\end{align}
where $b^\mu=0$ for YB-deformations of Minkowski spacetime, and $a^{\mu\nu\rho}$ 
and $b^\mu$ should be chosen such that $r$ satisfies the homogeneous CYBE. 
In this case, the $\beta$-field in the YB-deformed background becomes
\begin{align}
 \beta = 2\,\eta\,a^{\mu\nu\rho}\,x_\mu\,\partial_\nu\wedge \partial_\rho 
+ \eta\,b^\mu\,(z\,\partial_z + x^\nu\,\partial_\nu)\wedge \partial_\mu\,,
\end{align}
and this provides the constant $Q$-flux,
\begin{align}
 Q = \eta\,\bigl(2\,a_\mu{}^{\nu\rho}+\delta_\nu^{[\nu}\,b^{\rho]}\bigr)\, 
\rmd x^\mu\otimes\partial_\nu \wedge \partial_\rho 
 + \eta\,b^\mu\, \rmd z\otimes \partial_z\wedge \partial_\mu \,. 
\end{align}
By compactifying some of $x^\mu$ directions, the background becomes a $T$-fold. 
Importantly, as long as the $r$-matrix solves the homogeneous CYBE, the deformed background 
is a solution of DFT. Therefore, the YB deformation is a very systematic procedure to obtain 
solutions with $Q$-fluxes in DFT. Although we have considered YB deformations of Minkowski 
and $\AdS5\times\rmS^5$ backgrounds, it is applicable to more general cases such as 
$\AdS3\times \rmS^3\times \rmS^3\times\rmS^1$ solutions.

\medskip

On the other hand, let us remember that the GSE exhibits one isometry direction. 
This may suggest that they are effectively a 9-dimensional theory. 
In this respect, as it was denoted in \cite{Baguet:2016prz}, it is still an open problem 
what is the explicit relation, if any, between the GSE and the 9-dimensional gauged supergravities 
that involve the gauging of the trombone symmetry of type IIB supergravity 
\cite{Bergshoeff:2002nv,FernandezMelgarejo:2011wx}\footnote{As it occurs 
with GSE, gauged supergravities that are obtained by gauging the trombone symmetry 
or dimensional reduction on non-unimodular group manifolds cannot be derived 
from an action principle.}. If this were the case, then an additional question is in order. 
As the trombone symmetry is considered an accidental symmetry which is broken at higher order 
$\alpha'$-corrections \cite{Bergshoeff:2002nv}, it would be interesting to seek for 
the relation between type IIB supergravity and GSE, including $\alpha'$-corrections. 

\medskip 

It is also interesting to study the Poisson-Lie (PL) T-duality in our context. 
The $\eta$-deformation \cite{Delduc:2013qra,Delduc:2014kha,Arutyunov:2015qva,Arutyunov:2013ega}, 
which is an example of YB-deformations, is related to another integrable 
deformation called the $\lambda$-deformation \cite{Sfetsos:2013wia,Demulder:2015lva,
Hollowood:2014qma,Hollowood:2014rla,Borsato:2016ose} 
via the PL T-duality \cite{Vicedo:2015pna,Hoare:2015gda}. 
This relation has been further generalized by intriguing works 
\cite{Sfetsos:2015nya,Chervonyi:2016bfl}. 
Hence by generalizing our work to include the modified CYBE case, it should be possible to 
study the PL T-duality in our context. In fact, it is remarkable that 
the PL T-duality in DFT has been discussed in the recent work \cite{Hassler:2017yza} 
from another angle, the global structure of DFT \footnote{In relation to the global structure, 
the topology of DFT is discussed in \cite{Hassler:2016srl}.}, 
independently of a series of our works. As a matter of course, 
these directions meet up at some point. 

\medskip

In summary, we have shed light on a non-geometric aspect of YB deformation. 
Namely, using the formulas \eqref{YB-formula} and \eqref{eq:div-formula}, we have established a mapping between YB deformations and non-geometric backgrounds involving $Q$-fluxes. 
We hope that our result could be the starting point to delve into the relation between integrable deformations and non-geometric backgrounds.

\subsection*{Acknowledgments}

We are very grateful to Ursula Carow-Watamura, Yukio Kaneko, Domenico Orlando, Jeong-Hyuck Park, Shigeki Sugimoto, and Satoshi Watamura 
for valuable discussions. 
We appreciate useful discussions during the workshops ``Geometry, Duality and Strings'' 
at Yukawa Institute for Theoretical Physics and 
``Noncommutative Geometry, Duality and Quantum Gravity'' 
at Department of Physics, Kyoto University. 
J.J.F-M gratefully acknowledges the support of JSPS (Postdoctoral Fellowship) and Fundaci\'on S\'eneca/Universidad de Murcia (Programa Saavedra Fajardo).
The work of J.S.\ was supported by the Japan Society for the Promotion of Science (JSPS). 
The work of K.Y.\ was supported by the Supporting Program for Interaction-based Initiative 
Team Studies (SPIRITS) from Kyoto University and by a JSPS Grant-in-Aid for Scientific Research 
(C) No.\,15K05051. 
This work was also supported in part by the JSPS Japan-Russia Research Cooperative Program.

%
%

\appendix

\section{Generating GSE solutions with Penrose limits}
\label{sec:Penrose-limit}

In this appendix, we consider Penrose limit \cite{Penrose,Gueven:2000ru} of 
YB-deformed $\AdS5\times\rmS^5$ backgrounds and reproduce solutions of GSE 
studied in Section \ref{sec:non-geometry-Minkowski}. 
The R-R fluxes in the YB-deformed $\AdS5\times\rmS^5$ backgrounds may disappear under the Penrose limit. In that case, the resulting backgrounds become purely NS-NS solutions of GSE. 

\medskip

Penrose limit \cite{Penrose,Gueven:2000ru} is formulated for the standard supergravity. 
But, at least so far, there is no general argument on Penrose limit for the GSE case. 
Hence, it is quite non-trivial whether it can be extended to GSE or not. 
Here, we will not discuss a general theory of Penrose limit for GSE, 
but explain how to take a scaling of the extra vector $I$. 
The point here is that a YB-deformed background contains a deformations parameter 
and $I$ is proportional to it. 
Hence, there is a freedom to scale the deformation parameter 
in taking a Penrose limit. Without scaling the deformation parameter, 5D Minkowski spacetime 
is obtained as in the undeformed case. On the other hand, by taking an appropriate scaling of 
the deformation parameter, one can obtain a non-trivial solution of GSE with non-vanishing extra vector fields. 
We refer to the latter manner as {\it the modified Penrose limit}.
As a result, this modified Penrose limit may be regarded as a technique to generate solutions 
of GSE\footnote{Without any general argument, it is not ensured that the resulting background 
should satisfy the GSE. However, this point can be overcome by directly checking the GSE 
for the resulting background. As far as we have checked, it seems likely that 
this procedure works well. }. 

\subsection{Penrose limit of Poincar\'e $\AdS5$} 
\label{sec:Penrose-Ads5}

Let us first recall how to take a Penrose limit of the Poincar\'e metric of $\AdS5$\,. 

\medskip

The metric is given by 
\begin{align}
 \dd s^2 = \frac{r^2}{R^2}\,\bigl(-\dd t^2 + \dd \vec{x}^2\bigr) + R^2\,\frac{\dd r^2}{r^2}\,,
\end{align}
where $\vec{x}=(x^1,x^2,x^3)$\,. 

\medskip 

The first task is to determine a null geodesic. 
Here we are interested in a radial null geodesic described by 
\begin{align}
 \Bigl(\frac{\dd s}{\dd \tau}\Bigr)^2 
 = R^2\,\frac{\dot{r}^2}{r^2} + \frac{r^2}{R^2}\,(-\dot{t}^2) =0\,.
\label{null}
\end{align}
Here $\tau$ is an affine parameter and the symbol ``$\cdot$'' denotes a derivative 
in terms of $\tau$\,. 
From the energy conservation, we obtain that 
\begin{align}
 \frac{r^2}{R^2}\,\dot{t} \equiv E \quad (\mbox{constant})\,.
\end{align}
Hereafter, we will set $E=1$ by rescaling $\tau$\,. 
Then the equation \eqref{null} can be rewritten as 
\begin{align}
\dot{r}^2 =1\,. 
\end{align}
Hence, we will take a solution as 
\begin{align}
r = -\tau\,,
\end{align}
by adjusting an integration constant to be zero. 
Then $t$ can also be determined as follows: 
\begin{align}
 t = - \frac{R^2}{\tau}\,. 
\end{align}
As a result, the radial null geodesic is described as 
\begin{align}
t = \frac{R^2}{r}\,. \label{ng}
\end{align}

\medskip

Let us take a Penrose limit by employing the radial null geodesic \eqref{ng}. 
The first step is to introduce a new variable $\tilde{t}$ as a fluctuation around the null geodesic as 
\begin{align}
t = \frac{R^2}{r} - \tilde{t}\,.
\end{align}
Then, the metric of Poincar\'e $\AdS5$ is rewritten into the pp-wave form: 
\begin{align}
 \dd s^2 = -2\,\dd r\,\dd \tilde{t} - \frac{r^2}{R^2}\,\dd \tilde{t}^2 + \frac{r^2}{R^2}\,\dd \vec{x}^2\,.
\end{align}

\medskip

Next, by further transforming the coordinates as
\begin{align}
 \vec{x} = \frac{R}{r}\,\vec{y}\,, \qquad \tilde{t} = v - \frac{1}{2r}\,\vec{y}^2\,, 
\end{align}
the metric can be rewritten as 
\begin{align}
\dd s^2 = -2\,\dd r\, \dd v + \dd \vec{y}^2 + \cO\bigl(1/R^2\bigr)\,. 
\end{align}
Finally, by taking the $R \to \infty$ limit, the metric of 5D Minkowski spacetime is obtained. 

\subsection{Penrose limits of YB-deformed $\AdS5\times\rmS^5$}
\label{sec:Penrose-YB-deform}

Our aim here is to consider the modified Penrose limit of YB-deformed $\AdS5\times\rmS^5$ 
with classical $r$-matrices satisfying the homogeneous CYBE. 
In the following, we will focus upon two examples of non-unimodular classical $r$-matrices. 

\subsubsection*{\underline{Example 1) \ [solution of section \ref{sec:AdS-P0-D}]$\quad\overset{\text{Penrose limit}}{\longrightarrow}$\quad [solution of section \ref{sec:Minkowski-example1}]}}

The first example is a YB-deformed background associated with $r=\frac{1}{2}\,P_0\wedge D$, 
which was studied in section \ref{sec:AdS-P0-D}. 
To take a Penrose limit of the background \eqref{eq:AdS-P0-D}, let us rescale the fields as follows:
\begin{align}
\begin{alignedat}{2}
 \rmd s^2&~~\to~~ \rmd\tilde{s}^2 = R^2\,\rmd s^2\,,\qquad &
 B_2& ~~\to~~ \tilde{B}_2= R^2\,B_2\,,
\\
 F_3&~~\to~~ \tilde{F}_3= R^2\,F_3\,,&
 F_5&~~\to~~ \tilde{F}_5= R^4\,F_5\,.
\end{alignedat}
\end{align}
After performing a coordinate transformation for the radial direction,
\begin{align}
 z=\frac{R^2}{r}\,,
\end{align}
the radial null geodesic is given by 
\begin{align}
 x^0=\frac{R^2}{r}\label{pdng1}\,.
\end{align}
This expression coincides with the one \eqref{ng} even after performing the deformation. 

\medskip

As in the case of Poincar\'e $\AdS5$\,, a new variable $\tilde{t}$ is introduced 
as a fluctuation around the null geodesic \eqref{pdng1}:
\begin{align}
 x^0=\frac{R^2}{r}-\tilde{t}\,.
\end{align}
Let us perform a further coordinate transformation,
\begin{align}
 \rho=\frac{R}{r}\, p\,,\qquad
 \tilde{t}=v-\frac{p^2}{2r}\,.
\end{align}
If the $R\to \infty$ limit is taken naively, one can perform the usual Penrose limit, 
but it again leads to 5D Minkowski spacetime as in the case of the Poincar\'e $\AdS5$\,. 

\medskip 

It is interesting to add a modification to the usual process. 
That is to rescale the deformation parameter $\eta$ as well,
\begin{align}
 \eta=R^2\,\xi\,.
\end{align}
We refer to this modification as the modified Penrose limit. 

\medskip 

By taking the $R\to\infty$ limit and also the flat limit of the $\rmS^5$ part, 
we obtain the YB-deformed Minkowski background \eqref{flatpen} with the following identifications:
\begin{align}
 \{x^+,\,x^-,\,x,\,y,\,z,\,\eta\} \ ~~\longleftrightarrow~~ \ \{r,\, v,\, \rho\sin\theta \cos\phi,\, 
 \rho\sin\theta\sin\phi,\, z,\,\xi\} \,.
\end{align}
Remarkably, all of the R-R fluxes have vanished under this modified Penrose limit.

\subsubsection*{\underline{Example 2) \ [solution of section \ref{sec:AdS-Drinfeld--Jimbo}]$\quad\overset{\text{Penrose limit}}{\longrightarrow}$\quad [solution of section \ref{sec:Minkowski-example2}]}}

Let us next consider another YB-deformed background studied 
in Sec.\,\ref{sec:AdS-Drinfeld--Jimbo}.
To consider a Penrose limit of the background \eqref{eq:HvT}, 
let us rescale the fields as follows: 
\begin{align}
\begin{alignedat}{2}
 \rmd s^2& ~~\to~~ \rmd\tilde{s}= R^2\,\rmd s^2\,,\qquad&
 B_2& ~~\to~~ \tilde{B}_2= R^2\,B_2\,, 
\\
 F_3& ~~\to~~ \tilde{F}_3= R^2\,F_3\,,\qquad&
 F_5& ~~\to~~ \tilde{F}_5= R^4\,F_5\,.
\end{alignedat}
\end{align}
After performing a coordinate transformation,
\begin{align}
 z=\frac{R^2}{r}\,,
\end{align}
we obtain a radial null geodesic, which again takes the form,
\begin{align}
 x^0=\frac{R^2}{r}
\label{pdng}\,.
\end{align}

\medskip

Let us next introduce a new variable $\tilde{t}$ as a fluctuation around the null geodesic \eqref{pdng}:
\begin{align}
x^0=\frac{R^2}{r}-\tilde{t}\,.
\end{align}
Then, we perform a further coordinate transformation 
\begin{align}
 x^1=\frac{R}{r}\,z\,,\qquad 
 \rho=\frac{R}{r}\,p\,,\qquad
 \tilde{t}=v-\frac{p^2+z^2}{2r} \,.
\end{align}
As in the previous case, the deformation parameter is rescaled as
\begin{align}
 \eta= R^2\,\xi\,.
\end{align}
After taking the $R\to \infty$ limit, the resulting background is given by \eqref{eq:penHvT} 
with the following replacements:
\begin{align}
 \{r,\,v,\,p,\,\theta,\,\xi\}\to \{x^+,\,x^-,\,\sqrt{x^2+y^2}\,,\arctan(y/x)\,,\eta\}\,.
\end{align}
Note that all of the R-R fluxes have vanished again as in the previous example \eqref{flatpen}.

%
%

\small

\providecommand{\href}[2]{#2}\begingroup\raggedright\endgroup

\end{document}